\documentclass[preprint]{ptephy_om}

\preprintnumber{KYUSHU-HET-381} 

\usepackage{graphics}

\usepackage{amsmath} 
\usepackage{amsthm} 
\usepackage{url} 
\usepackage{stmaryrd}
\usepackage{tikz}
\usepackage{tikz-feynhand}
\usepackage{mathtools}
\usepackage{bm}
\usepackage{booktabs}
\usepackage{subcaption}
\usepackage{color}

\DeclareMathOperator{\tr}{tr}
\DeclareMathOperator{\Tr}{Tr}

\newcommand{\Slash}[1]{{\ooalign{\hfil/\hfil\crcr$#1$}}}
\numberwithin{equation}{section}

\allowdisplaybreaks

\begin{document}

\title{Wilsonian exact renormalization group flow in the gauged Nambu--Jona-Lasinio model}


\author[1]{Sorato Nagao}
\affil[1]{Department of Physics, Kyushu University, 744 Motooka, Nishi-ku,
Fukuoka 819-0395, Japan}

\author[1,2]{Hiroshi Suzuki}
\affil[2]{Quantum and Spacetime Research Institute (QuaSR), Kyushu University,
744 Motooka, Nishi-ku, Fukuoka 819-0395, Japan}





\begin{abstract}%
We study the Wilsonian exact renormalization group flow of the gauged
Nambu--Jona-Lasinio (NJL) model, a four-dimensional $U(1)$ gauge theory with
chirally invariant four-Fermi interactions, by using the gradient flow exact
renormalization group~(GFERG). Our gauge-invariant ansatz (truncation) for the
1PI Wilson action keeps a general photon kinetic
function~$\mathcal{K}_\tau(k^2)$, the fermion mass, and the scalar and vector
four-Fermi couplings. For a fixed gauge coupling~$e^2$, the Gaussian and NJL
(quasi-)fixed points on the critical surface merge at~$e_c^2\simeq17$, as
anticipated by the ladder Schwinger--Dyson equation and the conventional
Wilsonian renormalization group. We then let $e^2$ itself flow. Owing to the
manifest gauge invariance of the GFERG, the flow of~$e^2$ is governed solely by
the photon anomalous dimension~$\gamma_A$, which remains positive along both
branches up to the merging point. Hence, no fixed point with~$e^2\neq0$ exists.
Flows from a sufficiently large~$e^2$ in the ultraviolet (UV) reach the
chirally broken phase even for vanishing four-Fermi couplings, so the gauge
interaction does trigger chiral symmetry breaking. The merging point, however,
is not a fixed point, and the gauge coupling vanishes in any continuum limit
within our truncation.
\end{abstract}

\subjectindex{B02,B05,B31,B32}

\maketitle

\section{Introduction}
\label{sec:1}
Whether a sufficiently strong $U(1)$ gauge interaction dynamically and
spontaneously breaks chiral symmetry, and whether the critical point at which
this happens defines a nontrivial continuum theory, has been a long-standing
question in quantum field theory. In the ladder approximation of the
Schwinger--Dyson equation for the fermion self-energy with a non-running gauge
coupling, chiral symmetry is broken for~$\alpha=e^2/(4\pi)>\alpha_c=\pi/3$, and
the mass anomalous dimension at the critical coupling
is~$\gamma_m=1$~\cite{Maskawa:1974vs,Maskawa:1975hx,Fukuda:1976zb}. Since
$\gamma_m=1$ makes the four-Fermi operator $(\Bar{\psi}\psi)^2$ (approximately)
marginal, the four-Fermi interactions must be included in the analysis, which
leads to the gauged Nambu--Jona-Lasinio (NJL)
model~\cite{Nambu:1961tp,Nambu:1961fr,Kondo:1988qd}. It has been argued that
the critical line of the gauged NJL model, and in particular the point at which
the critical coupling of the four-Fermi interaction meets that of the gauge
interaction, provides a nontrivial fixed point and thus a nontrivial continuum
limit of strongly coupled quantum electrodynamics
(QED)~\cite{Miransky:1984ef,Bardeen:1985sm,Leung:1985sn}. Phenomenologically,
the large mass anomalous dimension was a key ingredient of walking
technicolor~\cite{Holdom:1984sk,Yamawaki:1985zg,Akiba:1985rr,Appelquist:1986an},
and the gauged NJL model was the basis of the top-quark
condensation~\cite{Miransky:1988xi,Miransky:1989ds,Nambu:1989jt,Marciano:1989xd,Bardeen:1989ds}; see~Ref.~\cite{Hill:2002ap} for a review. The question was
also addressed by lattice simulations of non-compact QED, which, however,
suggest that the theory is trivial, i.e., that the renormalized coupling
vanishes in the continuum limit~\cite{Gockeler:1997dn}.

In the language of the Wilsonian renormalization group (RG), i.e., the exact
(or functional) RG (ERG)~\cite{Wilson:1973jj,Morris:1993qb,Becchi:1996an,Pawlowski:2005xe,Igarashi:2009tj,Rosten:2010vm,Dupuis:2020fhh,Reuter:2019byg}, the
problem is the following. In the space of the gauge coupling~$e^2$ and the
four-Fermi couplings, if the running of~$e^2$ is neglected, there exist two
fixed points for small~$e^2$: the Gaussian one and the NJL one, the latter of
which separates the chirally symmetric and broken phases. As $e^2$ increases,
these two fixed points approach each other and merge at a critical
value~$e_c^2$; for~$e^2>e_c^2$, the RG flow generically goes to the chirally
broken phase~\cite{Aoki:1996fh}. This merging of fixed points is a typical
mechanism of the loss of conformality~\cite{Kaplan:2009kr}. The crucial
question is then what happens when the RG running of the gauge coupling itself,
which is caused by the vacuum polarization effects of the fermion, is taken
into account. An ERG study of QED with four-Fermi interactions has shown that
the gauge coupling grows toward the ultraviolet (UV) and the chirally broken
phase is reached before the Landau pole~\cite{Gies:2004hy}. On the other hand,
a recent ERG study of the gauged NJL model~\cite{Echigo:2025dia}, in which the
gauge coupling also runs, found a fixed point located very close to the merging
point. Since the conclusion hinges on a delicate balance between the running of
the gauge coupling and that of the four-Fermi couplings, it is desirable to
examine the problem within a framework in which the gauge symmetry is under
control and the approximation can be systematically improved.

The gradient flow exact renormalization
group~(GFERG)~\cite{Sonoda:2020vut,Sonoda:2022fmk,Sonoda:2025tyu} (see
also~Refs.~\cite{Sonoda:2019ibh,Miyakawa:2021hcx,Miyakawa:2021wus,Abe:2022smm,Miyakawa:2022qbz,Miyakawa:2023yob,Haruna:2023spq,Nagao:2025sri}) is a formulation
of the Wilsonian ERG that is suited to this purpose. In the GFERG, the
coarse-graining is defined by the gradient flow equations~\cite{Narayanan:2006rf,Luscher:2010iy,Luscher:2011bx,Luscher:2013cpa}, and, for the $U(1)$ gauge
theory, the resulting Wilson action satisfies a gauge Ward--Takahashi (WT)
identity which, in terms of the one-particle irreducible (1PI) Wilson action,
takes the conventional form~\cite{Sonoda:2022fmk}: the 1PI Wilson action is the
sum of a manifestly gauge-invariant functional and the standard gauge-fixing
term. This allows us to write down a manifestly gauge-invariant ansatz
(truncation) for the 1PI Wilson action, without the gauge non-invariant terms
such as the photon mass that are generated by a momentum cutoff in the
conventional ERG.\footnote{%
Here, it is the gauge covariance of the coarse-graining that matters; in this
respect, the GFERG is akin to the block-spin RG in lattice gauge
theory~\cite{Wilson:1979wp,Swendsen:1981rb} (see
also~Ref.~\cite{Carosso:2018bmz}), in which the coarse-graining is also defined
in a gauge-covariant way. In the GFERG, however, the coarse-graining is
continuous in the scale, and the theory is formulated in the continuum and
in the infinite volume from the outset.}
In~Refs.~\cite{Nagao:2026vaf,Nagao:2026gbq}, we derived and solved the GFERG
equation for a gauge-invariant ansatz for~QED that consists of the fermion and
the photon kinetic terms and the mass term, together with the minimal gauge
interaction. In this paper, we extend the analysis
of~Refs.~\cite{Nagao:2026vaf,Nagao:2026gbq} to the gauged NJL model, i.e., we
add the chirally invariant four-Fermi interactions to the ansatz, and study the
RG flow in the space of the gauge coupling~$e_\tau^2$, the four-Fermi couplings
$G_{S\tau}$ and~$G_{V\tau}$, the mass parameter~$m_\tau$, and the photon kinetic
function~$\mathcal{K}_\tau(k^2)$.

A characteristic feature of the GFERG equation is that it contains higher
functional derivatives of the variables of the 1PI Wilson action with respect
to those of the Wilson action; consequently, the right-hand side contains
two- and three-loop integrals, even though the flow equation is exact and the
ansatz is truncated. With the four-Fermi interactions, an explicit term-by-term
computation of these functional derivatives becomes impractical. We instead
evaluate them numerically as Taylor coefficients with respect to sources
placed at independent loop momenta, keeping the dependence on the couplings
and on~$\mathcal{K}_\tau(k^2)$ exact (Appendix~\ref{sec:A}).

Our findings are summarized as follows. If the running of the gauge coupling is
neglected (but the vacuum polarization is still included in~$\mathcal{K}_\tau$),
we reproduce the above picture: the Gaussian-type and the NJL-type quasi-fixed
points approach each other and merge at~$e_c^2\simeq16.8$ (in the Landau
gauge), where the mass anomalous dimension is~$\gamma_m\simeq1.43$. However,
the anomalous dimension of the gauge field~$\gamma_A$ is positive along both
branches of the quasi-fixed points up to the merging point. This implies that
the gauge coupling always decreases toward the infrared~(IR), i.e., increases
toward the UV, and no fixed point with~$e_\tau^2\neq0$ exists. Solving the full
RG flow including the running of~$e_\tau^2$ and of~$\mathcal{K}_\tau(k^2)$, we
confirm that the RG trajectories merely pass through the vicinity of the
merging point. A sufficiently strong gauge interaction does trigger chiral
symmetry breaking, but the merging point is not a fixed point, and the gauge
coupling vanishes in any continuum limit within our truncation. This conclusion
persists for the gauge-fixing parameter~$\xi_\tau=1$, $10$, and~$30$, although
the location of the merging point depends on~$\xi_\tau$.

This paper is organized as follows. In~Sect.~\ref{sec:2}, we recapitulate the
GFERG equation and introduce our ansatz for the gauged NJL model.
In~Sect.~\ref{sec:3}, we solve the resulting RG flow numerically and present
the results. Section~\ref{sec:4} is devoted to the conclusion. The numerical
method for evaluating the functional derivatives in the GFERG equation is
explained in~Appendix~\ref{sec:A}.

\section{Formulation}
\label{sec:2}
In this paper, we study the GFERG flow of the gauged NJL model, i.e., a
four-dimensional $U(1)$ gauge theory with a single fermion flavor ($N_f=1$)
and chirally invariant four-Fermi interactions. For the $U(1)$ gauge theory,
the GFERG flow equation for the 1PI Wilson action~$\mathit{\Gamma}_\tau$ is
given by
\begin{align}
   &\frac{\partial}{\partial\tau}
   \mathit{\Gamma}_\tau[\mathcal{A},\Bar{\Psi},\Psi]
\notag\\
   &=\int_p\,
   \biggl(
   \left[\frac{\delta}{\delta A_\mu(p)}
   +\frac{\delta\mathit{\Gamma}_\tau}{\delta\mathcal{A}_\mu(p)}\right]
   \left[
   \left(2p^2-\epsilon-\gamma_A\right)\mathcal{A}_\mu(p)
   +\frac{\delta\mathit{\Gamma}_\tau}{\delta\mathcal{A}_\mu(-p)}\right]
\notag\\
   &\qquad\qquad\qquad{}
   +\left(\frac{D+2}{2}+p\cdot\frac{\partial}{\partial p}\right)
   \left[\mathcal{A}_\mu(p)
   -\frac{\delta\mathit{\Gamma}_\tau}{\delta\mathcal{A}_\mu(-p)}\right]
   \cdot\frac{\delta\mathit{\Gamma}_\tau}{\delta\mathcal{A}_\mu(p)}
\notag\\
   &\qquad\qquad\qquad{}
   +\left[
   \frac{\delta}{\delta\Bar{\psi}(p)}
   +\frac{\delta}{\delta\Bar{\Psi}(p)}\mathit{\Gamma}_\tau
   \right](-2p^2+\gamma_\psi)
   \Bar{\Psi}(p)
\notag\\
   &\qquad\qquad\qquad{}
   +\left[
   \frac{\delta}{\delta\psi(p)}
   +\frac{\delta}{\delta\Psi(p)}\mathit{\Gamma}_\tau
   \right](-2p^2+\gamma_\psi)
   \Psi(p)
\notag\\
   &\qquad\qquad\qquad{}
   -i\left[
   \frac{\delta}{\delta\psi(p)}
   +\frac{\delta}{\delta\Psi(p)}\mathit{\Gamma}_\tau
   \right]
   \frac{\delta}{\delta\Bar{\Psi}(-p)}\mathit{\Gamma}_\tau
\notag\\
   &\qquad\qquad\qquad{}
   +\left(\frac{D+1}{2}+p\cdot\frac{\partial}{\partial p}\right)
   \left[
   \Psi(p)-i\frac{\delta}{\delta\Bar{\Psi}(-p)}\mathit{\Gamma}_\tau
   \right]
   \cdot\frac{\delta}{\delta\Psi(p)}\mathit{\Gamma}_\tau
\notag\\
   &\qquad\qquad\qquad{}
   +\left(\frac{D+1}{2}+p\cdot\frac{\partial}{\partial p}\right)
   \left[
   \Bar{\Psi}(p)+i\frac{\delta}{\delta\Psi(-p)}\mathit{\Gamma}_\tau
   \right]
   \cdot\frac{\delta}{\delta\Bar{\Psi}(p)}
   \mathit{\Gamma}_\tau
   \biggr)
\notag\\
   &\qquad{}
   +\int_{p,k}\,
   \left[
   \frac{\delta}{\delta\Bar{\psi}(p+k)}
   +\frac{\delta}{\delta\Bar{\Psi}(p+k)}\mathit{\Gamma}_\tau
   \right]
   (-4)p_\mu e_\tau\left[
   \mathcal{A}_\mu(k)
   +\frac{\delta}{\delta A_\mu(-k)}\right]
   \Bar{\Psi}(p)
\notag\\
   &\qquad{}
   +\int_{p,k}\,
   \left[
   \frac{\delta}{\delta\psi(p+k)}
   +\frac{\delta}{\delta\Psi(p+k)}\mathit{\Gamma}_\tau
   \right]
   4p_\mu e_\tau\left[
   \mathcal{A}_\mu(k)
   +\frac{\delta}{\delta A_\mu(-k)}\right]
   \Psi(p)
\notag\\
   &\qquad{}
   +\int_{p,k,l}\,
   \left[
   \frac{\delta}{\delta\Bar{\psi}(p+k+l)}
   +\frac{\delta}{\delta\Bar{\Psi}(p+k+l)}\mathit{\Gamma}_\tau
   \right]
   (-2)e_\tau^2\left[
   \mathcal{A}_\mu(k)
   +\frac{\delta}{\delta A_\mu(-k)}\right]
\notag\\
   &\qquad\qquad\qquad\qquad\qquad\qquad{}
   \times\left[
   \mathcal{A}_\mu(l)
   +\frac{\delta}{\delta A_\mu(-l)}\right]
   \Bar{\Psi}(p)
\notag\\
   &\qquad{}
   +\int_{p,k,l}\,
   \left[
   \frac{\delta}{\delta\psi(p+k+l)}
   +\frac{\delta}{\delta\Psi(p+k+l)}\mathit{\Gamma}_\tau
   \right]
   (-2)e_\tau^2\left[
   \mathcal{A}_\mu(k)
   +\frac{\delta}{\delta A_\mu(-k)}\right]
\notag\\
   &\qquad\qquad\qquad\qquad\qquad\qquad{}
   \times\left[
   \mathcal{A}_\mu(l)
   +\frac{\delta}{\delta A_\mu(-l)}\right]
   \Psi(p).
\label{eq:(2.1)}
\end{align}
In this expression, $\int_p:=\int d^Dp/(2\pi)^D$, $\epsilon:=(4-D)/2$ (we set
$D=4$ in what follows), $\tau$ parametrizes the scale logarithmically and
$\tau\to\infty$ is the IR direction; $\gamma_A$ and~$\gamma_\psi$ are the
anomalous dimensions of the gauge and fermion fields, respectively, and are
simply given by the logarithmic $\tau$-derivative of the wave function
renormalization factors~\cite{Nagao:2026vaf}.
Variables~$\{\mathcal{A}_\mu,\Psi,\Bar{\Psi}\}$ are the arguments of the 1PI
Wilson action, whereas $\{A_\mu,\psi,\Bar{\psi}\}$ are the arguments of the
(non-1PI) Wilson action. These two sets are related by the Legendre
transformation,
\begin{align}
   A_\mu(x)
   &=\mathcal{A}_\mu(x)
   -\frac{\delta\mathit{\Gamma}_\tau}{\delta\mathcal{A}_\mu(x)},&&
\notag\\
   \psi(x)&=\Psi(x)
   -i\frac{\delta}{\delta\Bar{\Psi}(x)}\mathit{\Gamma}_\tau,&
   \Bar{\psi}(x)&=\Bar{\Psi}(x)
   +i\frac{\delta}{\delta\Psi(x)}\mathit{\Gamma}_\tau.&
\label{eq:(2.2)}
\end{align}
From these relations, Eq.~\eqref{eq:(2.1)} can be written as
\begin{align}
   &\frac{\partial}{\partial\tau}
   \mathit{\Gamma}_\tau[\mathcal{A},\Bar{\Psi},\Psi]
\notag\\
   &=\int_p\,
   \biggl[
   \left(2p^2+\frac{D+2}{2}-\epsilon-\gamma_A
   +p\cdot\frac{\partial}{\partial p}\right)\mathcal{A}_\mu(p)
   \cdot\frac{\delta\mathit{\Gamma}_\tau}{\delta\mathcal{A}_\mu(p)}
\notag\\
   &\qquad\qquad{}
   +\left(
   2p^2+\frac{D+1}{2}-\gamma_\psi+p\cdot\frac{\partial}{\partial p}
   \right)
   \Bar{\Psi}(p)\cdot
   \frac{\delta}{\delta\Bar{\Psi}(p)}\mathit{\Gamma}_\tau
\notag\\
   &\qquad\qquad{}
   +\left(
   2p^2+\frac{D+1}{2}-\gamma_\psi+p\cdot\frac{\partial}{\partial p}
   \right)
   \Psi(p)\cdot
   \frac{\delta}{\delta\Psi(p)}\mathit{\Gamma}_\tau\biggr]
\notag\\
   &\qquad{}
   +\int_p\,
   \biggl[
   \left(2p^2-\epsilon-\gamma_A+1\right)
   \frac{\delta\mathcal{A}_\mu(p)}{\delta A_\mu(p)}
\notag\\
   &\qquad\qquad\qquad{}
   +\left(-2p^2+\gamma_\psi-\frac{1}{2}\right)
   \frac{\delta}{\delta\Bar{\psi}(p)}
   \Bar{\Psi}(p)
   +\left(-2p^2+\gamma_\psi-\frac{1}{2}\right)
   \frac{\delta}{\delta\psi(p)}
   \Psi(p)
   \biggr]
\notag\\
   &\qquad{}
   +\int_{p,k}\,
   (-4)e_\tau p_\mu\left[
   \frac{\delta}{\delta\Bar{\psi}(p+k)}
   +\frac{\delta}{\delta\Bar{\Psi}(p+k)}\mathit{\Gamma}_\tau
   \right]
   \left[
   \mathcal{A}_\mu(k)
   +\frac{\delta}{\delta A_\mu(-k)}\right]
   \Bar{\Psi}(p)
\notag\\
   &\qquad{}
   +\int_{p,k,l}\,
   (-2)e_\tau^2\left[
   \frac{\delta}{\delta\Bar{\psi}(p+k+l)}
   +\frac{\delta}{\delta\Bar{\Psi}(p+k+l)}\mathit{\Gamma}_\tau
   \right]
   \left[
   \mathcal{A}_\mu(k)
   +\frac{\delta}{\delta A_\mu(-k)}\right]
\notag\\
   &\qquad\qquad\qquad\qquad\qquad\qquad{}
   \times\left[
   \mathcal{A}_\mu(l)
   +\frac{\delta}{\delta A_\mu(-l)}\right]
   \Bar{\Psi}(p)
\notag\\
   &\qquad{}
   +\int_{p,k}\,
   4e_\tau p_\mu\left[
   \frac{\delta}{\delta\psi(p+k)}
   +\frac{\delta}{\delta\Psi(p+k)}\mathit{\Gamma}_\tau
   \right]
   \left[
   \mathcal{A}_\mu(k)
   +\frac{\delta}{\delta A_\mu(-k)}\right]
   \Psi(p)
\notag\\
   &\qquad{}
   +\int_{p,k,l}\,
   (-2)e_\tau^2\left[
   \frac{\delta}{\delta\psi(p+k+l)}
   +\frac{\delta}{\delta\Psi(p+k+l)}\mathit{\Gamma}_\tau
   \right]
   \left[
   \mathcal{A}_\mu(k)
   +\frac{\delta}{\delta A_\mu(-k)}\right]
\notag\\
   &\qquad\qquad\qquad\qquad\qquad\qquad{}
   \times\left[
   \mathcal{A}_\mu(l)
   +\frac{\delta}{\delta A_\mu(-l)}\right]
   \Psi(p),
\label{eq:(2.3)}
\end{align}
neglecting constant terms on the right-hand side; in the second integral, we
have used $\int_p\,f(p^2)\delta\Psi(p)/\delta\psi(p)=
\int_p\,f(p^2)\delta\Bar{\Psi}(p)/\delta\Bar{\psi}(p)$ for any function~$f$
(summed over the spinor index), which follows from the (graded) symmetry of the
second functional derivatives of~$\mathit{\Gamma}_\tau$.

A step-by-step derivation of the flow equation~\eqref{eq:(2.1)} is given
in~Ref.~\cite{Nagao:2026vaf}. One starts with the Wilson action including also
the Faddeev--Popov ghost sector but, for the $U(1)$ gauge theory, the Wilson
action for the ghost sector is completely decoupled and can be solved exactly.
The Becchi--Rouet--Stora--Tyutin (BRST) WT identity is then reduced to the
gauge WT identity in which only gauge and fermion fields appear. See
also~Refs.~\cite{Miyakawa:2021hcx,Miyakawa:2021wus,Sonoda:2022fmk}. The GFERG
equation~\eqref{eq:(2.1)} or~\eqref{eq:(2.3)} depends on the gauge
coupling~$e_\tau$ and contains functional derivatives of the 1PI variables with
respect to the variables of the Wilson action up to the third order, such
as~$\delta^3\Bar{\Psi}/(\delta\Bar{\psi}\delta A_\mu\delta A_\mu)$. These make
the required computation quite complicated, but this is the price for
preserving the manifest gauge symmetry. The conventional ERG (with an
appropriate momentum cutoff function) would be obtained by dropping the last
four terms of~Eq.~\eqref{eq:(2.3)}, which explicitly contain~$e_\tau$ and
originate from the gauge-covariant diffusion equation, the gradient flow
equation. Then, the expression contains only the first functional derivatives
which are given by the inverse of the matrix of the second functional
derivatives of the 1PI Wilson action (cf.~Eq.~(2.45)
of~Ref.~\cite{Nagao:2026vaf}); the flow equation takes the one-loop form
of the Wetterich equation~\cite{Nicoll:1977hi,Wetterich:1992yh,Wetterich:1993ne,Morris:1993qb}.

In the GFERG, it can be shown~\cite{Sonoda:2022fmk} (see
also~Ref.~\cite{Nagao:2026vaf}) that the $\tau$-evolution
by~Eq.~\eqref{eq:(2.1)} in conjunction with the RG flow equations for the gauge
coupling~$e_\tau$ and the gauge-fixing parameter~$\xi_\tau$,
\begin{equation}
   \frac{d}{d\tau}e_\tau^2=-2\gamma_Ae_\tau^2,\qquad
   \frac{d}{d\tau}\xi_\tau=2(\epsilon+\gamma_A)\xi_\tau,
\label{eq:(2.4)}
\end{equation}
preserves the following simple structure
\begin{equation}
   \mathit{\Gamma}_\tau[\mathcal{A},\Bar{\Psi},\Psi]
   =\mathit{\Gamma}_\tau^{\text{inv}}[\mathcal{A},\Bar{\Psi},\Psi]
   -\frac{1}{2\xi_\tau}\int_k\,
   e^{k^2}\mathcal{A}_\mu(-k)e^{k^2}\mathcal{A}_\nu(k)k_\mu k_\nu.
\label{eq:(2.5)}
\end{equation}
In this expression, the part~$\mathit{\Gamma}_\tau^{\text{inv}}$ is invariant
under the local $U(1)$ gauge transformation,
\begin{align}
   \delta\mathcal{A}_\mu(k)&=ik_\mu\chi(k),&&
\notag\\
   \delta\Psi(p)&=ie_\tau\int_k\,\chi(k)\Psi(p-k),&
   \delta\Bar{\Psi}(-p)
   &=-ie_\tau\int_k\,\chi(k)\Bar{\Psi}(-p-k).
\label{eq:(2.6)}
\end{align}
Note that the last, gauge non-invariant term in~Eq.~\eqref{eq:(2.5)} with the
gauge-fixing parameter~$\xi_\tau$ takes a fixed form. The
structure~\eqref{eq:(2.5)} with the RG equations~\eqref{eq:(2.4)} is a
consequence of the gauge WT identity that the GFERG equation~\eqref{eq:(2.1)}
preserves. In particular, the non-transverse part of the photon two-point
function is completely fixed by~Eq.~\eqref{eq:(2.5)}, and the flow of~$e_\tau$
is governed solely by the anomalous dimension~$\gamma_A$; these properties are
quite similar to those in the conventional 1PI effective action in
(perturbative) QED. Since the gauge transformation~\eqref{eq:(2.6)} takes the
\emph{conventional\/} form, it is straightforward to write down possible local
terms of~$\mathit{\Gamma}_\tau^{\text{inv}}$.

Now, since Eq.~\eqref{eq:(2.1)} is a nonlinear functional differential
equation, an exact solution in the infinite-dimensional theory space cannot be
expected, and some approximation is unavoidable. For a nonperturbative
analysis, in this paper, we study the RG flow of the following ansatz
(truncation) for the 1PI Wilson action,
\begin{align}
   \mathit{\Gamma}_\tau[\mathcal{A},\Bar{\Psi},\Psi]
   &=-\frac{1}{2}\int_k\,
   e^{k^2}\mathcal{A}_\mu(-k)e^{k^2}\mathcal{A}_\nu(k)
   \left[
   (k^2\delta_{\mu\nu}-k_\mu k_\nu)\mathcal{K}_\tau(k^2)
   +\frac{1}{\xi_\tau}k_\mu k_\nu
   \right]
\notag\\
   &\qquad{}
   -\int_p\,
   \Bar{\Psi}_{-1}(-p)(\Slash{p}+im_\tau)\Psi_{-1}(p)
   +e_\tau\int_{p,k}\,
   \Bar{\Psi}_{-1}(-p-k)e^{k^2}\Slash{\mathcal{A}}(k)\Psi_{-1}(p)
\notag\\
   &\qquad{}
   -\int d^Dx\,
   \frac{G_{S\tau}}{2}\left\{
   \left[\Bar{\Psi}_{-1}(x)\Psi_{-1}(x)\right]^2
   -\left[\Bar{\Psi}_{-1}(x)\gamma_5\Psi_{-1}(x)\right]^2
   \right\}
\notag\\
   &\qquad{}
   -\int d^Dx\,
   \frac{G_{V\tau}}{2}\left\{
   \left[\Bar{\Psi}_{-1}(x)\gamma_\mu\Psi_{-1}(x)\right]^2
   +\left[\Bar{\Psi}_{-1}(x)\gamma_5\gamma_\mu\Psi_{-1}(x)\right]^2
   \right\},
\label{eq:(2.7)}
\end{align}
where $m_\tau$ is the fermion mass parameter, and $G_{S\tau}$ and~$G_{V\tau}$ are
the scalar-type and the vector-type four-Fermi couplings, respectively. The
function~$\mathcal{K}_\tau(k^2)$ in the first line describes a nontrivial
evolution of the transverse part of the gauge field action. In this paper, we
normalize the gauge potential by imposing
\begin{equation}
   \mathcal{K}_\tau(k^2=0)=1.
\label{eq:(2.8)}
\end{equation}
The analyticity of the function~$\mathcal{K}_\tau(k^2)$, i.e., the locality, on
the other hand, implies in particular
\begin{equation}
   \left.k^2\mathcal{K}_\tau'(k^2)\right|_{k^2=0}=0.
\label{eq:(2.9)}
\end{equation}

Equation~\eqref{eq:(2.7)} may be regarded as the simplest extension, by
four-Fermi interactions, of the truncation for the 1PI Wilson action in QED
whose RG flow has been fully studied
in~Refs.~\cite{Nagao:2026vaf,Nagao:2026gbq}. In~Eq.~\eqref{eq:(2.7)}, field
variables with the subscript~$-1$, the ``$-1$
variables''~\cite{Sonoda:2022fmk}, are defined from the original variables
through the solution to the flow equations (written in coordinate space)
\begin{align}
   \partial_t\mathcal{A}_{t\mu}(x)
   &=\partial_\nu
   \left[
   \partial_\nu\mathcal{A}_{t\mu}(x)-\partial_\mu\mathcal{A}_{t\nu}(x)
   \right]
   +\alpha_0\partial_\mu\partial_\nu\mathcal{A}_{t\nu}(x),&
   \mathcal{A}_{0\mu}(x)&=\mathcal{A}_\mu(x),
\notag\\
   \partial_t\Psi_t(x)
   &=\left\{
   \left[\partial_\mu-ie\mathcal{A}_{t\mu}(x)\right]^2
   +ie\alpha_0\left[\partial_\mu\mathcal{A}_{t\mu}(x)\right]
   \right\}\Psi_t(x),&
   \Psi_0(x)&=\Psi(x),
\notag\\
   \partial_t\Bar{\Psi}_t(x)
   &=\left\{
   \left[\partial_\mu+ie\mathcal{A}_{t\mu}(x)\right]^2
   -ie\alpha_0\left[\partial_\mu\mathcal{A}_{t\mu}(x)\right]
   \right\}\Bar{\Psi}_t(x),&
   \Bar{\Psi}_0(x)&=\Bar{\Psi}(x),
\label{eq:(2.10)}
\end{align}
by solving \emph{backward\/} from~$t=0$ to~$t=-1$. Equation~\eqref{eq:(2.10)}
is nothing but the gradient flow equations
in~Refs.~\cite{Luscher:2010iy,Luscher:2011bx,Luscher:2013cpa} specialized to
the $U(1)$ gauge theory. In~Eq.~\eqref{eq:(2.10)}, the parameter~$e$ is
regarded as a fixed constant and, after solving the equations from~$t=0$
to~$t=-1$, $e$~is replaced by~$e_\tau$. For~$\alpha_0=1$, which we set
throughout this paper, $\mathcal{A}_{-1\mu}(x)$ is simply given by
\begin{equation}
   \mathcal{A}_{-1\mu}(x)=e^{-\partial^2}\mathcal{A}_\mu(x)
   \Rightarrow
   \mathcal{A}_{-1\mu}(k)=e^{k^2}\mathcal{A}_\mu(k),
\label{eq:(2.11)}
\end{equation}
whereas $\Psi_{-1}(x)$ and~$\Bar{\Psi}_{-1}(x)$ are of infinite order
in~$\mathcal{A}_\mu(x)$ through Eq.~\eqref{eq:(2.10)}. See Appendix~B
of~Ref.~\cite{Nagao:2026gbq} for a detailed study. It can also be seen that,
for the gauge transformation~\eqref{eq:(2.6)}, defining the gauge
transformation function~$\chi_t(k)$ by
\begin{equation}
   \partial_t\chi_t(k)=-\alpha_0k^2\chi_t(k),\qquad
   \chi_0(k)=\chi(k),
\label{eq:(2.12)}
\end{equation}
the $-1$~variables transform as
\begin{align}
   \delta\mathcal{A}_{-1\mu}(k)&=ik_\mu\chi_{-1}(k),&&
\notag\\
   \delta\Psi_{-1}(p)&=ie_\tau\int_k\,\chi_{-1}(k)\Psi_{-1}(p-k),&
   \delta\Bar{\Psi}_{-1}(-p)
   &=-ie_\tau\int_k\,\chi_{-1}(k)\Bar{\Psi}_{-1}(-p-k).
\label{eq:(2.13)}
\end{align}
Therefore, the ansatz~\eqref{eq:(2.7)} is, apart from the gauge-fixing term
whose form is determined by~Eq.~\eqref{eq:(2.5)}, invariant under the gauge
transformation~\eqref{eq:(2.6)}.

We use the $-1$~variables in our ansatz~\eqref{eq:(2.7)} because the
ansatz~\eqref{eq:(2.7)} reduces to the Gaussian fixed point of the GFERG
for~$G_{S\tau}=G_{V\tau}=0$, $e_\tau\to0$, and~$\mathcal{K}_\tau\to1$, as shown
in~Ref.~\cite{Sonoda:2022fmk}. We regard this as a minimal requirement for the
ansatz of the 1PI Wilson action.\footnote{%
Since $\Bar{\Psi}(x)\Psi(x)$ is also gauge invariant, we could use the original
variables~$\{\Psi,\Bar{\Psi}\}$ for the four-Fermi interactions as well. Here,
we use $\{\Psi_{-1},\Bar{\Psi}_{-1}\}$ so that all fermion fields in the ansatz
appear through the same variables; as a consequence, the four-Fermi vertices
also couple to the gauge potential through the $-1$~variables.} A simple
gauge-invariant ansatz in terms of the original
variables~$\{\mathcal{A}_\mu,\Psi,\Bar{\Psi}\}$ would not necessarily contain
the Gaussian fixed point in a simple form.

In the GFERG, chiral symmetry is expressed by a WT identity, which reduces to
the Ginsparg--Wilson relation~\cite{Ginsparg:1981bj} when the Wilson action is
quadratic in the fermion fields~\cite{Miyakawa:2021hcx,Sonoda:2022fmk}. Since
this identity is nonlinear in the 1PI Wilson action, it is not easy to
construct an ansatz that possesses exact chiral symmetry. Our
ansatz~\eqref{eq:(2.7)} is not chirally symmetric either, even for~$m_\tau=0$;
the mass parameter~$m_\tau$ receives an additive renormalization from both the
gauge interaction and the four-Fermi interactions, and we include this effect
numerically in the RG flow below.

\section{Nonperturbative RG flow}
\label{sec:3}
What we basically do is to substitute the ansatz~\eqref{eq:(2.7)} into the
GFERG equation~\eqref{eq:(2.1)} or~\eqref{eq:(2.3)}, equate both sides to
obtain differential equations for the parameters $m_\tau$, $G_{S\tau}$,
and~$G_{V\tau}$ and the function~$\mathcal{K}_\tau(k^2)$ in the ansatz, and solve
them numerically together with~Eq.~\eqref{eq:(2.4)}.\footnote{%
The anomalous dimensions $\gamma_A$ and~$\gamma_\psi$ are also determined by
the equality~\eqref{eq:(2.1)}, together with the normalization
condition~\eqref{eq:(2.8)} and the normalization of the fermion kinetic term
in~Eq.~\eqref{eq:(2.7)}, respectively.} Without the four-Fermi interactions,
i.e., for~$G_{S\tau}=G_{V\tau}=0$, this has been carried out
in~Refs.~\cite{Nagao:2026vaf,Nagao:2026gbq}. With the four-Fermi interactions,
however, an explicit computation of the functional derivatives
in~Eq.~\eqref{eq:(2.1)} or~\eqref{eq:(2.3)} becomes an extremely painful task.
In this paper, instead of computing these functional derivatives analytically,
we evaluate them numerically as Taylor coefficients with respect to sources
placed at independent loop momenta; the dependence on the couplings and on the
function~$\mathcal{K}_\tau(k^2)$ is kept exactly. The details are given
in~Appendix~\ref{sec:A}.

In what follows, we mainly work in the Landau gauge, $\xi_\tau=0$, which is
preserved under~Eq.~\eqref{eq:(2.4)}. Unfortunately, it turns out that, with
our ansatz~\eqref{eq:(2.7)}, the coefficients of the flow equations, such as
the anomalous dimension~$\gamma_A$ and the additive renormalization of the mass
parameter~$m_\tau$, depend on the gauge-fixing parameter~$\xi_\tau$; see the
Note added in the fourth arXiv version~(v4) of~Ref.~\cite{Nagao:2026vaf}. A
detailed analysis of the $\xi_\tau$-dependence of the Wilson action will be
given elsewhere~\cite{Nagao:2026xxx}. At the end of this section, we will
comment on how our observations depend on the value of~$\xi_\tau$.

\begin{figure}[htbp]
\centering
\includegraphics[width=0.9\textwidth]{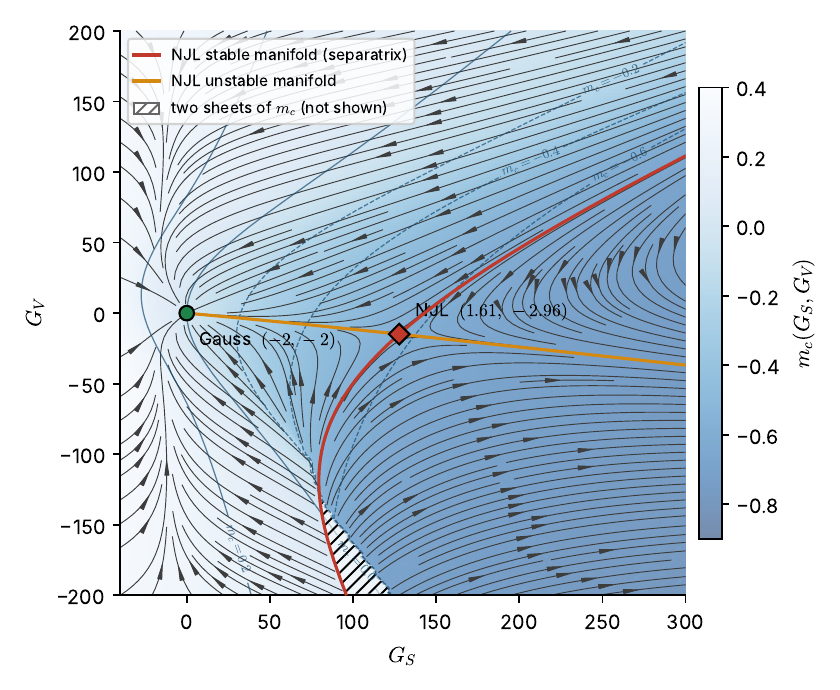}
\caption{The RG flow in the $(G_S,G_V)$~plane for~$e_\tau^2=0$, i.e., the NJL
model; arrows indicate the direction of increasing~$\tau$ (toward the IR). For
each $(G_S,G_V)$, the critical value of the mass parameter~$m_c(G_S,G_V)$ (the
color map) is numerically determined so that $m_\tau$ does not run away under
the RG flow; we regard the surface $m=m_c(G_S,G_V)$ as the critical surface
corresponding to the massless fermion, and the flow on it is shown. The green
circle indicates the Gaussian fixed point and the red diamond the NJL fixed
point, which has one relevant direction within the critical surface; the
numbers in parentheses are the eigenvalues of the linearized flow on the
critical surface. The orange curve is the unstable manifold of the NJL fixed
point, i.e., the renormalized trajectory associated with it. The red curve is
the stable manifold of the NJL fixed point, which separates the chirally
symmetric and broken phases. In the hatched region, the critical surface has
two sheets and the flow is not shown.}
\label{fig:1}
\end{figure}

Let us start with the case without the gauge interaction, $e_\tau^2=0$. This
subspace is preserved under the RG flow~\eqref{eq:(2.4)} and the system simply
becomes the NJL model~\cite{Nambu:1961tp,Nambu:1961fr}. The RG flow in the
$(G_S,G_V)$~plane for~$e_\tau^2=0$ is depicted in~Fig.~\ref{fig:1}.
We see that there are two fixed points: the Gaussian one and a nontrivial NJL
one. The latter, which corresponds to the critical coupling separating the
chirally symmetric and broken phases, has long been known from the large-$N_f$
gap equation and ERG studies of four-Fermi
theories~\cite{Aoki:1996fh,Gies:2004hy}. See also
Ref.~\cite{Cresswell-Hogg:2026jpe} for a recent argument that the four-Fermi
theory in four dimensions is renormalizable.

\begin{figure}[htbp]
\centering
\includegraphics[width=0.9\textwidth]{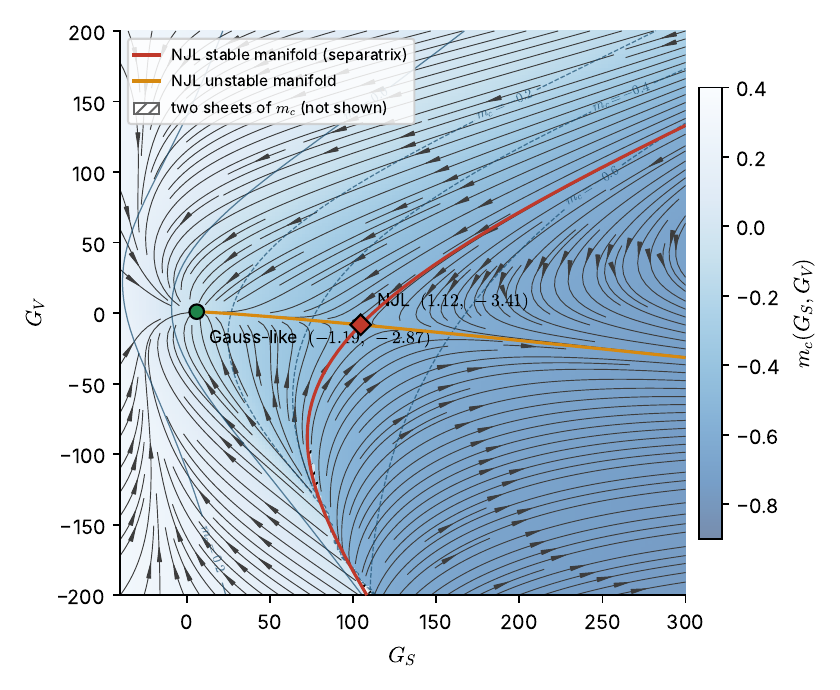}
\caption{The same as~Fig.~\ref{fig:1}, but for a fixed (frozen) gauge
coupling~$e_\tau^2=10$ (i.e., the running of~$e_\tau^2$ is neglected). The full
right-hand side of the GFERG equation~\eqref{eq:(2.1)} is included, and the
approximate stationary solution of~$\mathcal{K}_\tau(k^2)$ is used (see the
text). The green circle and the red diamond indicate the Gaussian-type and the
NJL-type quasi-fixed points, respectively.}
\label{fig:2}
\end{figure}

\begin{figure}[htbp]
\centering
\includegraphics[width=0.9\textwidth]{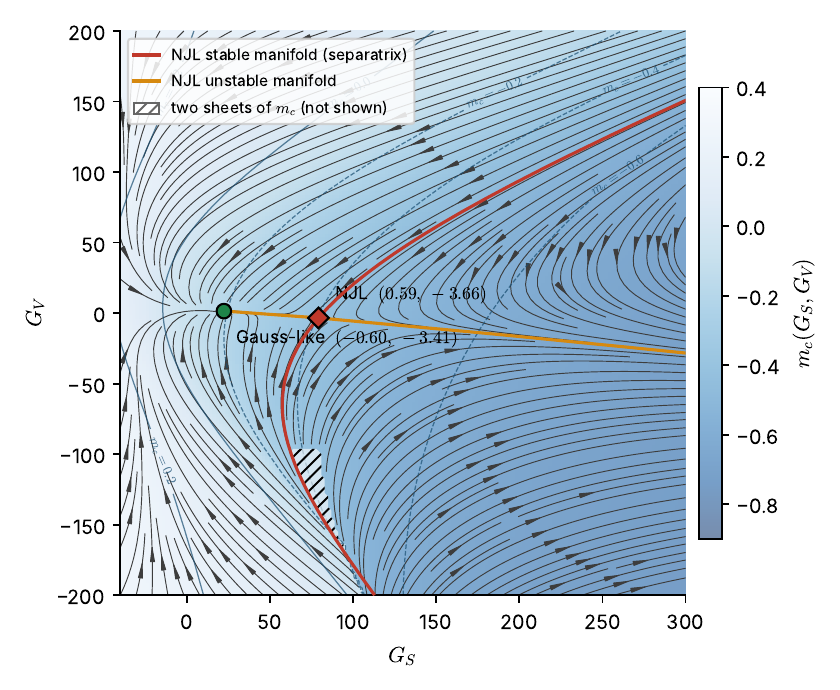}
\caption{The same as~Fig.~\ref{fig:2}, but for~$e_\tau^2=15$.}
\label{fig:3}
\end{figure}

\begin{figure}[htbp]
\centering
\includegraphics[width=0.9\textwidth]{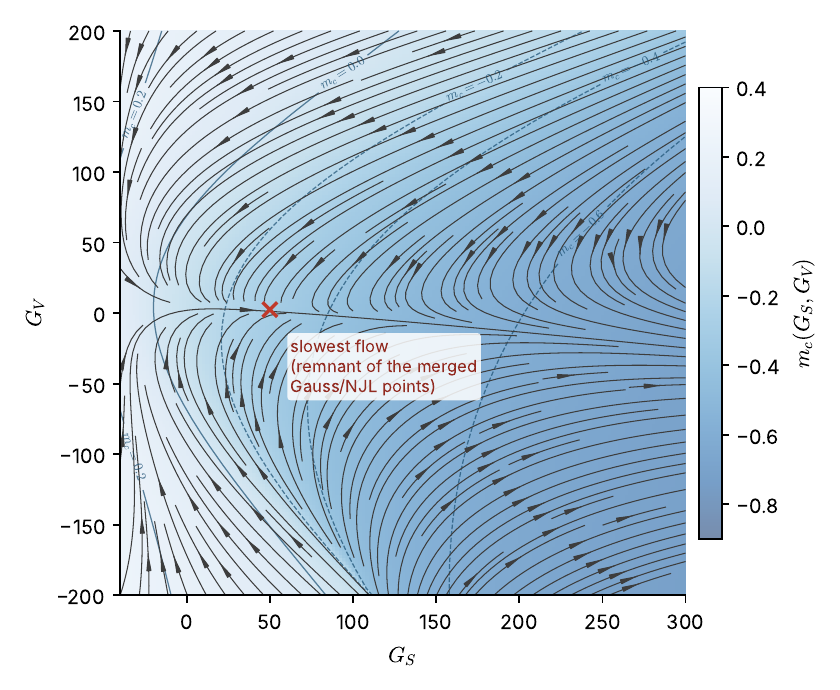}
\caption{The same as~Fig.~\ref{fig:2}, but for~$e_\tau^2=20>e_c^2\simeq16.8$. No
quasi-fixed point exists for this value of~$e_\tau^2$; the cross indicates the
point at which the flow is slowest, a remnant of the merged quasi-fixed
points.}
\label{fig:4}
\end{figure}

Next, we turn the gauge interaction on. First, we neglect the RG running of
the gauge coupling and \emph{treat $e_\tau^2$ as a fixed (or frozen)
parameter}. This treatment corresponds to that in~Refs.~\cite{Maskawa:1974vs,Maskawa:1975hx,Fukuda:1976zb,Kondo:1988qd,Aoki:1996fh}, in which the radiative
corrections to the photon propagator are neglected and the gauge coupling does
not run. In our analysis, however, the vacuum polarization effect is still
taken into account through the photon kinetic function~$\mathcal{K}_\tau(k^2)$:
we use the stationary solution of its flow equation at fixed~$e_\tau^2$,
$\partial_\tau\mathcal{K}_\tau(k^2)=0$, determined self-consistently at each
quasi-fixed point. For the flow in the $(G_S,G_V)$~plane
in~Figs.~\ref{fig:2}--\ref{fig:4}, to reduce the computational cost, a
stationary $\mathcal{K}_\tau(k^2)$ at~$G_S=G_V=0$ is employed for each value
of~$m_\tau$. Its dependence on $G_S$ and~$G_V$ is less than~$0.5\%$; we also
find that the flow with a crude approximation~$\mathcal{K}_\tau(k^2)=1$ shows
qualitatively the same result.

Figures~\ref{fig:2} and~\ref{fig:3} show the RG flow for~$e_\tau^2=10$
and~$e_\tau^2=15$, respectively.
Compared with~Fig.~\ref{fig:1}, we see that the two quasi-fixed points approach
each other as $e_\tau^2$ increases: the Gaussian-type point moves
from~$(G_S,G_V)=(0,0)$ to~$(6.0,1.0)$ and~$(22.4,1.4)$, and the NJL-type point
from~$(127.8,-14.9)$ to~$(104.6,-8.3)$ and~$(79.2,-3.4)$, for~$e_\tau^2=10$
and~$15$, respectively. These are only quasi-fixed points, because the gauge
coupling~$e_\tau^2$ actually runs at those points, $\gamma_A\neq0$, as we will
see below.

The two quasi-fixed points merge at~$e_c^2\simeq16.8$
(at~$(G_S,G_V)\simeq(49,0.05)$), and no quasi-fixed point exists
for~$e_\tau^2>e_c^2$ (see~Fig.~\ref{fig:4} for~$e_\tau^2=20$).

The merging of the two quasi-fixed points at the critical value of the gauge
coupling~$e_c^2$ is interesting. Since for~$e_\tau^2>e_c^2$ the RG flow
generically leads to~$G_{S\tau}\to+\infty$, i.e., to the chirally broken phase,
it has sometimes been suggested that chiral symmetry is dynamically and
spontaneously broken in strongly coupled
QED~\cite{Maskawa:1974vs,Maskawa:1975hx,Kondo:1988qd,Aoki:1996fh}. It has even
been argued that the merging point provides a nontrivial fixed point and thus a
nontrivial continuum limit with broken chiral
symmetry~\cite{Miransky:1984ef,Bardeen:1985sm,Leung:1985sn}. A recent ERG study
of the gauged NJL model~\cite{Echigo:2025dia}, in which the gauge coupling also
runs, found a fixed point located very close to the merging point, which
appears to support this possibility.

In what follows, we solve the RG flow of the ansatz~\eqref{eq:(2.7)} without
the above simplification, i.e., including the running of the gauge
coupling~$e_\tau^2$ and of the photon kinetic function~$\mathcal{K}_\tau(k^2)$
(in the Landau gauge, $\xi_\tau=0$, for the time being). First of all, we note
that Eq.~\eqref{eq:(2.4)} implies that $\gamma_A=0$ at any fixed point
with~$e_\tau^2\neq0$; conversely, if $\gamma_A>0$, $e_\tau^2$ monotonically
decreases toward the IR.

To see whether such a fixed point exists, we compute~$\gamma_A$ along the two
branches of quasi-fixed points obtained above, as a function of~$e_\tau^2$;
$\gamma_A$ is determined by imposing the normalization
condition~\eqref{eq:(2.8)}. In this calculation, the stationary photon kinetic
function~$\mathcal{K}_\tau(k^2)$ is determined self-consistently at each
quasi-fixed point, including also the effect of $G_{S\tau}$ and~$G_{V\tau}$. We
note that the four-Fermi interactions do not enter the right-hand side of the
GFERG equation for the photon two-point function directly. Thus $\gamma_A$ and
the flow of~$\mathcal{K}_\tau(k^2)$ depend on $G_{S\tau}$ and~$G_{V\tau}$ only
indirectly, through $m_\tau$ and~$\gamma_\psi$ (see Appendix~\ref{sec:A.1}). The
result is shown in~Fig.~\ref{fig:5}.
\begin{figure}[htbp]
\centering
\includegraphics[width=\textwidth]{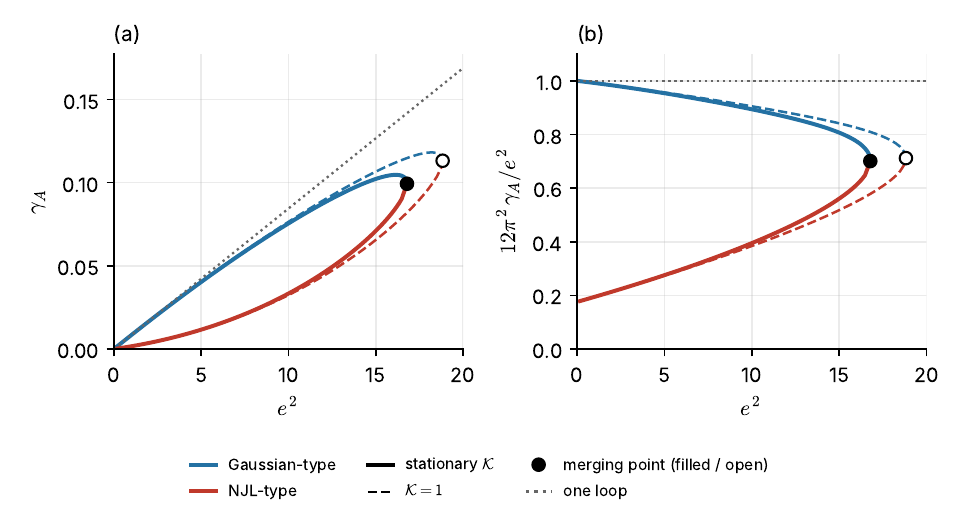}
\caption{(a)~The anomalous dimension~$\gamma_A$ along the two branches of
quasi-fixed points as a function of the fixed~$e_\tau^2$; the blue and red
curves are for the Gaussian-type and the NJL-type quasi-fixed points,
respectively. The solid curves are obtained with the stationary solution
of~$\mathcal{K}_\tau(k^2)$, and the dashed curves with the crude
approximation~$\mathcal{K}_\tau(k^2)=1$ for reference. The filled (open) dot
indicates the merging point for the former (latter). The dotted line is the
one-loop value~$e_\tau^2/(12\pi^2)$. (b)~The same quantity normalized by the
one-loop value.}
\label{fig:5}
\end{figure}
We see that $\gamma_A$ remains positive along both branches up to the merging
point, where $\gamma_A\simeq0.099$ ($0.113$ with~$\mathcal{K}_\tau=1$). Hence,
no fixed point with~$e_\tau^2\neq0$ exists.

\begin{figure}[htbp]
\centering
\includegraphics[width=0.9\textwidth]{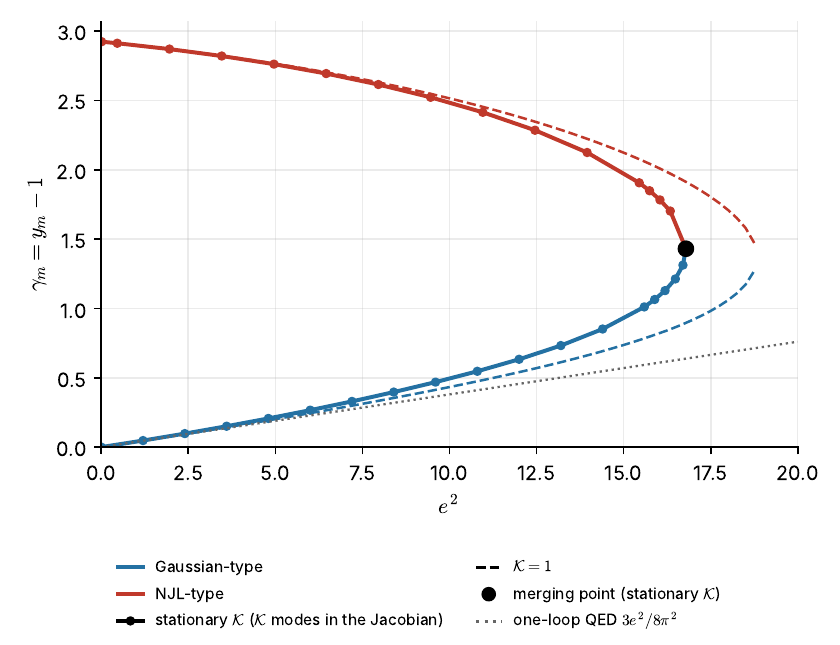}
\caption{The mass anomalous dimension~$\gamma_m=y_m-1$ along the two branches
of quasi-fixed points as a function of the fixed~$e_\tau^2$; the blue and red
curves are for the Gaussian-type and the NJL-type quasi-fixed points,
respectively. The solid curves (with dots at the computed points) are obtained
with the stationary solution of~$\mathcal{K}_\tau(k^2)$, whose fluctuations are
included in the stability matrix; the dashed curves are with the crude
approximation~$\mathcal{K}_\tau(k^2)=1$. The black dot indicates the merging
point, and the dotted line is the one-loop value~$3e_\tau^2/(8\pi^2)$.}
\label{fig:6}
\end{figure}

We also examine the mass anomalous dimension~$\gamma_m:=y_m-1$, where $y_m$ is
the eigenvalue of the linearized flow at the quasi-fixed point whose
eigenvector is dominated by the mass direction. In computing~$y_m$, we include
the fluctuations of~$\mathcal{K}_\tau(k^2)$ in the stability matrix, together
with those of~$m$, $G_S$, and~$G_V$; freezing $\mathcal{K}_\tau$ changes
$\gamma_m$ by less than~$10^{-2}$. See~Fig.~\ref{fig:6}.
Along the Gaussian-type branch, $\gamma_m$ increases monotonically from its
one-loop value~$3e_\tau^2/(8\pi^2)$ at small~$e_\tau^2$, reaches $\gamma_m=1$
at~$e_\tau^2\simeq15.5$, and takes the value (in the Landau gauge)
\begin{equation}
   \gamma_m\simeq1.43
\label{eq:(3.1)}
\end{equation}
at the merging point~$e_c^2\simeq16.8$ (with~$\mathcal{K}_\tau=1$, we obtain
$\gamma_m\simeq1.37$ at~$e_c^2\simeq18.8$). Along the NJL-type branch,
$\gamma_m$ decreases from~$\gamma_m\simeq2.92$ at~$e_\tau^2=0$ toward the same
value at the merging point. Since the two branches meet with a square-root
singularity, $\gamma_m$ varies rapidly near~$e_c^2$; for instance,
$\gamma_m\simeq1.31$ on the Gaussian-type branch at~$e_\tau^2=16.70$. The value
at the merging point thus exceeds $\gamma_m=1$ obtained at the critical
coupling in the ladder Schwinger--Dyson approach, although such a comparison is
subject to the scheme dependence of both $e_c^2$ and~$\gamma_m$ away from a
true fixed point.

\begin{figure}[htbp]
\centering
\includegraphics[width=0.9\textwidth]{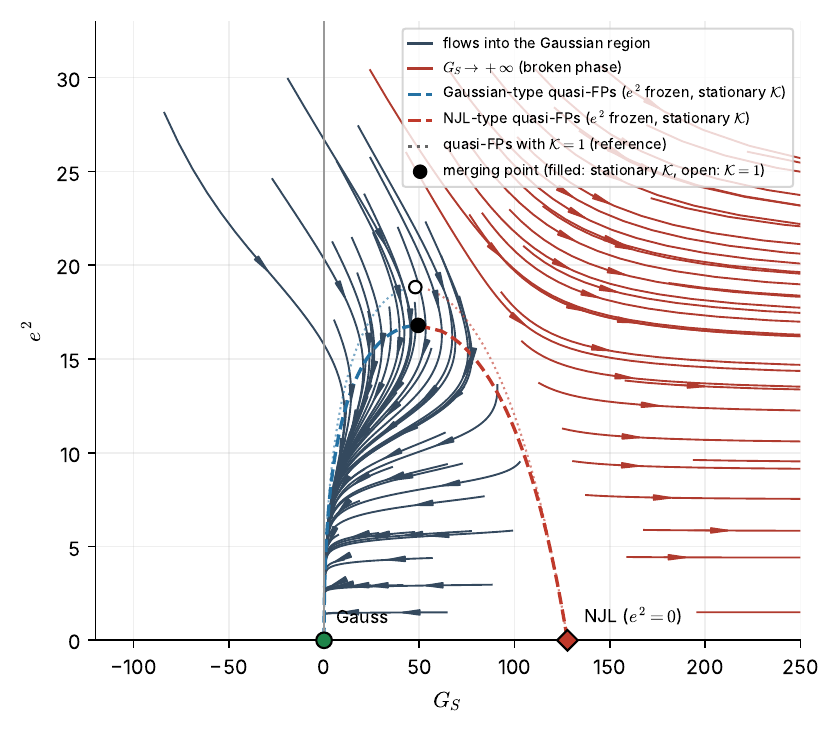}
\caption{The RG flow on the surface of renormalized trajectories
$\{m=m_c,G_V=G_V^c\}$, projected onto the $(G_S,e^2)$~plane, with the full
flow of~$\mathcal{K}_\tau(k^2)$ and the running of~$e_\tau^2$ included (in the
Landau gauge). The initial transient~$\tau<0.8$, during which $G_V$
and~$\mathcal{K}_\tau$ relax to the surface, is not shown. The dark curves flow
into the Gaussian fixed point (green circle), and the red curves flow
to~$G_S\to+\infty$ (chirally broken phase). The dashed curves are the
Gaussian-type (blue) and the NJL-type (red) quasi-fixed points with
frozen~$e_\tau^2$ and the stationary~$\mathcal{K}_\tau$ (Fig.~\ref{fig:5}), and
the dotted curves the same with~$\mathcal{K}_\tau=1$; the filled (open) dot is
the corresponding merging point. The red diamond is the NJL fixed point
at~$e_\tau^2=0$.}
\label{fig:7}
\end{figure}

To show the absence of any true fixed point around the merging point more
directly, we solve the RG flow including all the effects, i.e., the running
of~$e_\tau^2$ and the full flow of~$\mathcal{K}_\tau(k^2)$. For each initial
value~$(e_0^2,G_{S0})$, with~$\mathcal{K}_\tau=1$ initially and~$G_{V0}$ chosen
close to the surface, the initial mass~$m_0$ is tuned by bisection so that
$m_\tau$ does not run away, and $G_V$ and~$\mathcal{K}_\tau$, being irrelevant,
relax to a two-dimensional surface; we refer to this surface as the surface of
renormalized trajectories and show the flow projected onto the
$(G_S,e^2)$~plane in~Fig.~\ref{fig:7}.
The gauge coupling~$e_\tau^2$ decreases monotonically along every trajectory,
and no fixed point appears in the region~$e_\tau^2>0$. The branch of NJL-type
quasi-fixed points approximately acts as a separatrix: trajectories on its
left flow into the Gaussian fixed point, while those on its right flow
to~$G_{S\tau}\to+\infty$, i.e., to the chirally broken phase. Trajectories
passing near the merging point do not stay there; it is merely a point that the
flow passes through. Note that a trajectory starting from a sufficiently large
$e_0^2$ reaches the broken phase even for a vanishing initial four-Fermi
coupling: for~$G_{S0}=0$, the flow goes to the Gaussian fixed point
for~$e_0^2\lesssim35$, whereas it goes to the broken phase for~$e_0^2\gtrsim40$.
In this sense, a sufficiently strong gauge interaction does trigger chiral
symmetry breaking. Since $e_\tau^2$ increases toward the UV, however, the
merging point is not a fixed point, and the gauge coupling vanishes in any
continuum limit within our truncation. This is the main conclusion of this
paper.

\begin{figure}[htbp]
\centering
\includegraphics[width=\textwidth]{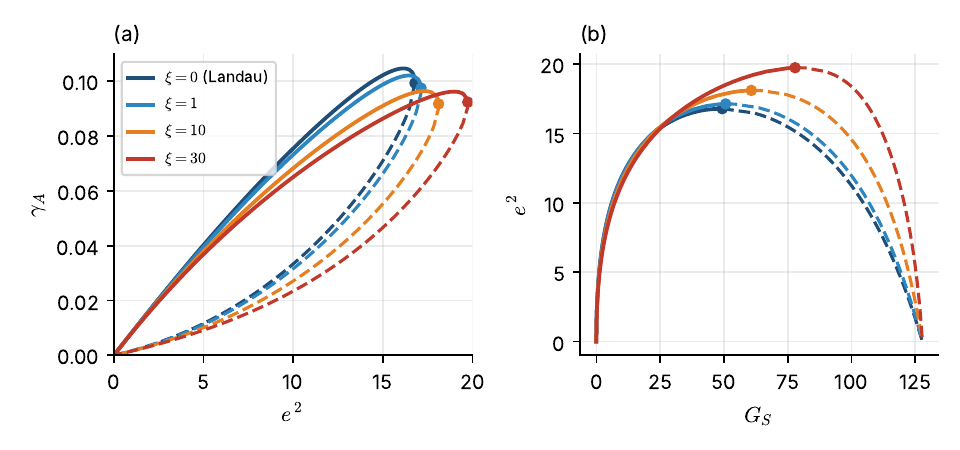}
\caption{Dependence on the gauge-fixing parameter. (a)~The anomalous
dimension~$\gamma_A$ along the branches of quasi-fixed points for~$\xi_\tau=0$
(Landau gauge), $1$, $10$, and~$30$ (different colors); the solid and dashed
curves are for the Gaussian-type and the NJL-type branches, respectively, and
the dots indicate the merging points. (b)~The same branches in the
$(G_S,e^2)$~plane. The stationary solution of~$\mathcal{K}_\tau(k^2)$ is used,
and the $\xi_\tau$-dependence is included up to two loops (see the text).}
\label{fig:8}
\end{figure}

Finally, we comment on the dependence on the gauge-fixing parameter. Repeating
the above analysis for~$\xi_\tau=1$, $10$, and~$30$ (Fig.~\ref{fig:8}), we find
that the merging point moves to~$e_c^2\simeq17.1$, $18.1$, and~$19.7$,
respectively, while $\gamma_A$ remains positive along both branches of the
quasi-fixed points up to the merging point ($\gamma_A\simeq0.09$--$0.10$ at the
merging point). In this analysis, the $\xi_\tau$-dependence of the right-hand
side is included up to two loops; we have checked that the $\xi_\tau$-dependent
three-loop contributions are smaller by one to two orders of magnitude. Hence,
although the precise location of the merging point depends on~$\xi_\tau$, our
conclusion that no fixed point with~$e_\tau^2\neq0$ exists is unchanged within
this range of~$\xi_\tau$. For much larger~$\xi_\tau$, the Gaussian-type branch
extends to the region of large~$G_S$ and positive~$m_\tau$, which is beyond
the range of our present numerical calculation.

\clearpage

\section{Conclusion}
\label{sec:4}
In this paper, we have studied the RG flow of the gauged NJL model, i.e.,
four-dimensional QED with~$N_f=1$ and the chirally invariant four-Fermi
interactions, by using the GFERG. Thanks to the gauge WT identity of the
conventional form, which the GFERG equation preserves, we could adopt a
manifestly gauge-invariant ansatz for the 1PI Wilson action, consisting of the
fermion and photon kinetic terms, the mass term, and the four-Fermi terms with
scalar--pseudoscalar and vector--axial-vector couplings. This is in contrast to
the ERG with a momentum cutoff, in which the gauge symmetry is expressed by
modified WT (or Slavnov--Taylor) identities and a gauge-consistent truncation
requires a careful construction~\cite{Bonini:1994kp,Ellwanger:1995qf,Echigo:2025dia}. The functional derivatives in the GFERG equation, which give rise to up
to three-loop integrals on the right-hand side, were evaluated numerically as
Taylor coefficients with respect to sources, with the dependence on the
couplings and on the photon kinetic function~$\mathcal{K}_\tau(k^2)$ kept
exactly.

Our main observations are as follows. If the running of the gauge coupling is
neglected, the Gaussian-type and the NJL-type quasi-fixed points approach each
other as $e_\tau^2$ increases and merge at~$e_c^2\simeq16.8$ (in the Landau
gauge), where the mass anomalous dimension is~$\gamma_m\simeq1.43$. This
reproduces, at least qualitatively, the picture based on the ladder
Schwinger--Dyson equation and the Wilsonian ERG with a frozen gauge coupling.
However, the anomalous dimension of the gauge field~$\gamma_A$ is positive
along both branches of the quasi-fixed points up to the merging point; thus,
the gauge coupling always decreases toward the IR and there is no fixed point
with~$e_\tau^2\neq0$. The full RG flow (Fig.~\ref{fig:7}), including the running
of~$e_\tau^2$ and of~$\mathcal{K}_\tau(k^2)$, confirms that the trajectories pass
through the vicinity of the merging point without staying there. A sufficiently
strong gauge interaction does drive the system into the chirally broken phase,
even from a vanishing initial four-Fermi coupling. The merging point, however,
is not a fixed point, and within our truncation the gauge coupling vanishes in
any continuum limit. In this sense, the scenario in which strongly coupled QED
possesses a nontrivial continuum limit with dynamically broken chiral symmetry
is not supported by our analysis. This conclusion holds also for the
gauge-fixing parameter~$\xi_\tau=1$, $10$, and~$30$, although the location of
the merging point depends on~$\xi_\tau$. It is consistent with lattice
simulations of non-compact QED~\cite{Gockeler:1997dn} and with the ERG
analysis of~Ref.~\cite{Gies:2004hy}, while it differs from the fixed point
found in~Ref.~\cite{Echigo:2025dia}. The difference might be attributed to how
the running of the gauge coupling near the merging point is treated; it would
be interesting to clarify this point. We also note that a different possibility
of a UV completion of QED, a fixed point with a nonvanishing Pauli coupling and
a vanishing gauge coupling, has been proposed
in~Refs.~\cite{Gies:2020xuh,Gies:2022aiz}. Since the Pauli term explicitly
breaks chiral symmetry, this possibility lies outside the present scope of our
chirally invariant ansatz.

Several limitations of the present analysis should be mentioned. First, our
ansatz is a truncation. In particular, the four-Fermi couplings are
momentum-independent, and higher-dimensional operators, such as the operator
outside the ansatz generated by the RG flow (the $O_X$ structure
in~Sect.~\ref{sec:A.4}) and multi-fermion operators, are neglected. Since the
merging point lies at a rather large coupling, $G_S\simeq49$, the effect of
these operators may not be negligible quantitatively, and the numerical values
such as~$e_c^2$ and~$\gamma_m$ should be taken with care. What is required for
the conclusion is the sign of~$\gamma_A$, which is positive with a margin
($\gamma_A\simeq0.1$ at the merging point) and is dominated by the fermion
vacuum polarization; we expect it to be robust against such improvements.
Second, as noted in~Sect.~\ref{sec:3}, the coefficients of the flow equations
depend on the gauge-fixing parameter~$\xi_\tau$ with our ansatz. Although our
conclusion is unchanged for~$\xi_\tau\leq30$, a systematic understanding of the
$\xi_\tau$-dependence is needed~\cite{Nagao:2026xxx}. Third, we have considered
only~$N_f=1$. For larger~$N_f$, the vacuum polarization becomes stronger and
$\gamma_A$ increases, so that we expect the same conclusion; nevertheless, the
flavor structure of the four-Fermi interactions becomes richer and should be
examined.

The numerical method developed in this paper does not rely on the specific form
of the ansatz: once the 1PI Wilson action is given as a functional of the
fields, the right-hand side of the GFERG equation can be evaluated without
deriving the explicit form of the functional derivatives. It can therefore be
applied to improved ans\"atze, including momentum-dependent four-Fermi
couplings and vertices, as have been developed in ERG studies of
QCD~\cite{Ellwanger:1995qf,Ellwanger:1996wy,Bergerhoff:1997cv,Mitter:2014wpa,Cyrol:2016tym,Cyrol:2017ewj,Corell:2018yil,Fu:2025hcm}, higher-dimensional
operators, and also to other gauge theories, such as the large-$N_f$ QED and,
possibly, non-Abelian gauge theories, which are relevant to the conformal
window. We hope to report on these in the near future.

\section*{Acknowledgments}
We would like to thank Hidenori Sonoda for many years of valuable discussions.
The work of H.S. was supported by JSPS KAKENHI Grant Number JP23739313 and by
the MEXT program ``Supporting Pioneering Research through AI for 1,000
Discovery Challenges (SPReAD),'' Grant Number JPMXP1726257479. It was also
partially supported by the Quantum and Spacetime Research Institute, Kyushu
University.
We received substantial assistance from Claude (Anthropic) and ChatGPT (OpenAI)
in carrying out this study. Needless to say, any remaining errors are our own.

\appendix

\section{Numerical evaluation of the functional derivatives}
\label{sec:A}
In this appendix, we explain how the right-hand side of the GFERG
equation~\eqref{eq:(2.3)} is evaluated for the ansatz~\eqref{eq:(2.7)}. The
basic idea is to regard the multiple functional derivatives with respect to the
variables of the (non-1PI) Wilson action as Taylor coefficients with respect to
nilpotent sources, and to obtain them by solving the Legendre
transformation~\eqref{eq:(2.2)} order by order in the sources. We never write
down the individual terms of the functional derivatives, which become extremely
numerous with the four-Fermi interactions; nevertheless, the dependence on the
couplings and on the function~$\mathcal{K}_\tau(k^2)$ is kept exactly.

\subsection{Structure of the right-hand side}
\label{sec:A.1}
Let us collectively denote the variables of the Wilson action
by~$\phi=(A_\mu,\Bar{\psi},\psi)$ and those of the 1PI Wilson action
by~$\Phi=(\mathcal{A}_\mu,\Bar{\Psi},\Psi)$. Equation~\eqref{eq:(2.2)} is then
written as
\begin{equation}
   \phi_a=\Phi_a-\sigma_{ab}\frac{\partial\mathit{\Gamma}_\tau}{\partial\Phi_b},
\label{eq:(A1)}
\end{equation}
where $\sigma$ is a constant matrix that pairs each component of~$\phi$ with
the derivative with respect to its conjugate variable, with entries~$1$
and~$\pm i$. The right-hand side of~Eq.~\eqref{eq:(2.3)} contains functional
derivatives of functions of~$\Phi$ with respect to~$\phi$. The first
derivative, $J_{ab}:=\partial\Phi_b/\partial\phi_a$, is the inverse of the
matrix of the second functional derivatives of~$\mathit{\Gamma}_\tau$, i.e., the
full propagator in the presence of the background fields; in fact,
Eq.~\eqref{eq:(A1)} implies
$J_{ac}[\delta_{cb}-\sigma_{bd}\partial^2\mathit{\Gamma}_\tau/
(\partial\Phi_c\partial\Phi_d)]=\delta_{ab}$. The higher derivatives then follow
from the chain rule. In this notation, the right-hand side
of~Eq.~\eqref{eq:(2.3)} takes the form
\begin{equation}
   \partial_\tau\mathit{\Gamma}_\tau
   =\mathcal{L}\mathit{\Gamma}_\tau
   +\mathcal{N}[\mathit{\Gamma}_\tau],
\label{eq:(A2)}
\end{equation}
where $\mathcal{L}$ is a linear first-order differential operator with respect
to~$\Phi$, and $\mathcal{N}$ consists of the terms containing at least one
derivative with respect to~$\phi$. In a condensed notation, we have
\begin{align}
   \mathcal{N}[\mathit{\Gamma}]
   &=\Tr\left[(K^A+1)J^{A\mathcal{A}}\right]
   +\Tr\left[(K^F-1/2)J^{\Bar{\psi}\Bar{\Psi}}\right]
   +\Tr\left[(K^F-1/2)J^{\psi\Psi}\right]
\notag\\
   &\quad{}
   +\Bar{C}_{ab}^i\left[
   \left(\frac{\partial}{\partial\Bar{\psi}_a}+\mathit{\Gamma}_{\Bar{\Psi}_a}
   \right)
   \left(\mathcal{A}_i+\frac{\partial}{\partial A_i}\right)\Bar{\Psi}_b
   -\mathit{\Gamma}_{\Bar{\Psi}_a}\mathcal{A}_i\Bar{\Psi}_b\right]
\notag\\
   &\quad{}
   +\Bar{Q}_{ab}^{ij}\left[
   \left(\frac{\partial}{\partial\Bar{\psi}_a}
   +\mathit{\Gamma}_{\Bar{\Psi}_a}\right)
   \left(\mathcal{A}_i+\frac{\partial}{\partial A_i}\right)
   \left(\mathcal{A}_j+\frac{\partial}{\partial A_j}\right)\Bar{\Psi}_b
   -\mathit{\Gamma}_{\Bar{\Psi}_a}\mathcal{A}_i\mathcal{A}_j\Bar{\Psi}_b
   \right]
\notag\\
   &\qquad{}
   +(\Bar{\psi}\to\psi,\Bar{\Psi}\to\Psi),
\label{eq:(A3)}
\end{align}
where $K^A:=2p^2-\epsilon-\gamma_A$, $K^F:=-2p^2+\gamma_\psi$,
$\mathit{\Gamma}_{\Bar{\Psi}_a}:=\partial\mathit{\Gamma}/\partial\Bar{\Psi}_a$,
and $\Bar{C}\sim-4e_\tau p_\mu$ and~$\Bar{Q}\sim-2e_\tau^2$ denote the
coefficients of the last four terms of~Eq.~\eqref{eq:(2.3)}, which originate
from the gauge-covariant diffusion equation. The trace terms in the first line
of~Eq.~\eqref{eq:(A3)} give one-loop contributions, the $\Bar{C}$~terms
contribute up to two loops, and the $\Bar{Q}$~terms up to three loops. Thus,
the right-hand side of the GFERG equation contains at most three-loop
integrals; we emphasize that this is an exact property of the equation, not a
result of the truncation.

For the $\Bar{\Psi}$-type terms \emph{with no background photon\/}, the
multi-loop contributions consist of the following six structures:
\begin{align}
   C_2&=-4e_\tau p_\mu
   \frac{\delta^2\Bar{\Psi}(p)}{\delta\Bar{\psi}\delta A_\mu},&
   Q_c&=-2e_\tau^2\frac{\delta^2\mathcal{A}_\mu}{\delta\Bar{\psi}\delta A_\mu}
   \Bar{\Psi}_{\text{bg}},&
   P_1&=-4e_\tau^2\frac{\delta\mathcal{A}_\mu}{\delta\Bar{\psi}}
   \frac{\delta\Bar{\Psi}}{\delta A_\mu},
\notag\\
   P_2&=-2e_\tau^2\frac{\delta\mathcal{A}_\mu}{\delta A_\mu}
   \frac{\delta\Bar{\Psi}}{\delta\Bar{\psi}},&
   Q_b&=-2e_\tau^2\mathit{\Gamma}_{\Bar{\Psi}}
   \frac{\delta^2\Bar{\Psi}}{\delta A_\mu\delta A_\mu},&
   Q_a&=-2e_\tau^2\frac{\delta^3\Bar{\Psi}}
   {\delta\Bar{\psi}\delta A_\mu\delta A_\mu}.
\label{eq:(A4)}
\end{align}
Among these, $C_2$, $Q_c$, $P_1$, $P_2$, and~$Q_b$ are two-loop contributions
and $Q_a$ is a three-loop one. The $\Psi$-type terms are related to the
$\Bar{\Psi}$-type ones by charge conjugation (see Appendix~B
of~Ref.~\cite{Nagao:2026vaf}); we have checked that they coincide after the
projection described in~Sect.~\ref{sec:A.4} (to~$10^{-14}$), and thus we compute
only the $\Bar{\Psi}$-type terms and multiply them by~$2$.

Note that the four-Fermi interactions do not contribute to the photon two-point
function. With no background fermion fields, the four-Fermi terms contribute to
a functional derivative of~$\mathit{\Gamma}_\tau$ only when it is taken with
respect to four fermion fields. On the other hand, by the chain rule, the $n$th
derivative~$\delta^n\Phi/\delta\phi^n$ is expressed in terms of the derivatives
of~$\mathit{\Gamma}_\tau$ up to the $(n+1)$th order, and the $(n+1)$th
derivative is taken with respect to the fields of the same types as those
in~$\delta^n\Phi/\delta\phi^n$, because $J$ does not mix the photon and the
fermion fields in the absence of background fermion fields. Since the
functional derivatives in~Eq.~\eqref{eq:(2.3)} are at most of the third order,
the four-Fermi terms could contribute only through the fourth derivative
of~$\mathit{\Gamma}_\tau$
in~$\delta^3\Bar{\Psi}/(\delta\Bar{\psi}\delta A_\mu\delta A_\mu)$ and its $\Psi$
counterpart. This fourth derivative, however, is taken with respect to two
fermion and two photon fields.

We thus evaluate the photon two-point function by using Eq.~(4.3)
of~Ref.~\cite{Nagao:2026vaf} (or, equivalently, Eq.~(A7)
of~Ref.~\cite{Nagao:2026gbq}), and the source inversion explained below is
needed only for the fermion bilinear and four-fermion parts. Accordingly, the
background fields in what follows consist of fermion fields only.

\subsection{Source inversion}
\label{sec:A.2}
To evaluate a derivative such
as~$\delta^nf(\Phi)/\delta\phi_{a_1}\cdots\delta\phi_{a_n}$ at the background, we
set
\begin{equation}
   \phi=\phi_{\text{bg}}+\sum_{s=1}^n\varepsilon_s,
\label{eq:(A5)}
\end{equation}
where each source~$\varepsilon_s$ is a single component of a field (with a
polarization vector for the photon, or with a spinor index and a Grassmann
generator~$\theta$ for the fermion) carrying an independent loop
momentum~$r_s$; we call it a ``slot.'' The derivative is then given by the
coefficient of~$\varepsilon_1\cdots\varepsilon_n$ in~$f(\Phi(\phi))$. Since
each slot appears at most once, we may regard the sources as nilpotent,
$\varepsilon_s^2=0$.

Now, splitting
$\mathit{\Gamma}_\tau=\mathit{\Gamma}_0+\mathit{\Gamma}_{\text{int}}$ into the free
part (the $\mathcal{A}^2$ and~$\Bar{\Psi}\Psi$ terms) and the interaction part,
we can rewrite Eq.~\eqref{eq:(A1)} as
\begin{equation}
   \Phi=J_0\left(
   \phi+\sigma\frac{\partial\mathit{\Gamma}_{\text{int}}}{\partial\Phi}
   \right),
\label{eq:(A6)}
\end{equation}
where the free propagator is given by (cf.~Eq.~(3.16)
of~Ref.~\cite{Nagao:2026vaf})
\begin{equation}
   J_0(p)=e^{-2p^2}\mathop{\text{diag}}
   \left(h_{\mu\nu}(p),\frac{1}{i}h_F(-p),\frac{1}{i}{}^th_F(p)\right),
\label{eq:(A7)}
\end{equation}
with
\begin{align}
   &h_{\mu\nu}(p)=\left(\delta_{\mu\nu}-\frac{p_\mu p_\nu}{p^2}\right)h_T(p^2)
   +\frac{p_\mu p_\nu}{p^2}h_L(p^2),
\notag\\
   &h_T(x):=\frac{1}{e^{-2x}+x\mathcal{K}_\tau(x)},\qquad
   h_L(x):=\frac{\xi_\tau}{\xi_\tau e^{-2x}+x},
\notag\\
   &h_F(p)=\frac{i}{e^{-2p^2}+i(\Slash{p}+im_\tau)}.
\label{eq:(A8)}
\end{align}

We call a subset~$I$ of the slots a ``sector,'' and denote by~$\delta\Phi_I$
the component of~$\Phi$ proportional to~$\prod_{s\in I}\varepsilon_s$. The
background fields are chosen according to the part to be extracted. For the
four-fermion part, we take $\Bar{\Psi}_{\text{bg}}=\Bar{\Psi}_\alpha(0)$
and~$\Psi_{\text{bg}}=\Psi_\alpha(0)$ at zero momentum. For the fermion bilinear
part, on the other hand, we need the dependence on the external momentum to
read off the coefficient of~$\Slash{p}$ (see~Sect.~\ref{sec:A.4}), and we take
$\Bar{\Psi}_{\text{bg}}(-p)$ and~$\Psi_{\text{bg}}(p)$ with~$p\neq0$. The
coefficient~$\delta\Phi_I$ is then a polynomial in these background Grassmann
generators. Since each slot carries a fixed momentum, $\delta\Phi_I$ in
momentum space is supported at the momenta
\begin{equation}
   \delta\Phi_I(q)
   =\sum_{c=-1}^1\delta(q-r_I-cp)\delta\Phi_{I,c},
   \qquad r_I=\sum_{s\in I}r_s,
\label{eq:(A9)}
\end{equation}
where $c$ is the number of~$\Psi_{\text{bg}}$ minus that of~$\Bar{\Psi}_{\text{bg}}$
contained in each term. Since we need only the terms bilinear in the background
fields for the fermion bilinear part, we may discard the terms containing two
or more factors of~$\Psi_{\text{bg}}$ or of~$\Bar{\Psi}_{\text{bg}}$, which cannot
contribute to the final result; hence $c\in\{-1,0,1\}$. For the four-fermion
part, $p=0$ and this decomposition is unnecessary.

Expanding Eq.~\eqref{eq:(A6)} in the sources, we have
\begin{equation}
   \delta\Phi_I
   =J_0(r_I)
   \left(\varepsilon_I
   +\sigma
   \left[\frac{\partial\mathit{\Gamma}_{\text{int}}}{\partial\Phi(-r_I)}\right]_I
   \right),
   \qquad
   \varepsilon_I
   =\begin{cases}
   \varepsilon_s&I=\{s\},\\0&|I|\geq2,
   \end{cases}
\label{eq:(A10)}
\end{equation}
where $[\cdots]_I$ denotes the part that contains each slot in~$I$ exactly
once. This part consists of two types of terms. The terms of the first type are
products of~$\delta\Phi_{I'}$ with proper subsets~$I'\subset I$ and the
background fields; these are known if we solve Eq.~\eqref{eq:(A10)} in the
order~$|I|=1$, $2$, \dots, $n$. The terms of the second type are linear
in~$\delta\Phi_I$ itself and take the form~$\sigma B\delta\Phi_I$, where
$B$~is a linear operator built from~$\mathit{\Gamma}_{\text{int}}$ and the
background fields. Since $\mathit{\Gamma}_{\text{int}}$ is at least cubic in the
fields and the background consists of fermion fields only, every term of~$B$
contains at least one background Grassmann generator; that is, $B$~is
nilpotent. Equation~\eqref{eq:(A10)} is thus solved as
\begin{equation}
   \delta\Phi_I=\Tilde{J}_0(\varepsilon_I+\sigma[\cdots]_I'),
\label{eq:(A11)}
\end{equation}
where $[\cdots]_I'$ denotes the terms without~$\delta\Phi_I$, and
\begin{equation}
   \Tilde{J}_0=(1-J_0\sigma B)^{-1}J_0=\sum_{k=0}^N(J_0\sigma B)^kJ_0
\label{eq:(A12)}
\end{equation}
is the propagator dressed by the background fields. The sum
in~Eq.~\eqref{eq:(A12)} terminates exactly, where $N$~is the number of
background Grassmann generators retained ($N=2$ and~$4$ for the fermion
bilinear and four-fermion parts, respectively). This is equivalent to summing
the geometric series~(3.15) of~Ref.~\cite{Nagao:2026vaf} to the required order.

In practice, we determine $\Tilde{J}_0$ column by column. Note that
$\Tilde{J}_0$ is a $12\times12$ matrix (four photon polarizations and four
spinor components each of~$\Bar{\psi}$ and~$\psi$), whose entries are
polynomials in the background generators, and that it depends on the sector
only through its momentum. For a single slot, $I=\{s\}$, we have
$[\cdots]_I'=0$, and Eq.~\eqref{eq:(A11)} reduces
to~$\delta\Phi_I=\Tilde{J}_0\varepsilon_s$. Solving this with~$\varepsilon_s$
set to each of the $12$ unit vectors of the field components, we obtain the
$12$ columns of~$\Tilde{J}_0$ at that momentum. For the four-fermion part, we
compute $\Tilde{J}_0$ in this way once at each sector momentum and use it for
all the source polarizations, spinor components, and structures
in~Eq.~\eqref{eq:(A4)} evaluated at the same loop momenta. For the fermion
bilinear part, on the other hand, $N=2$ and $\Tilde{J}_0$ connects the
different momenta~$r_I+cp$; it is then less costly to solve
Eq.~\eqref{eq:(A10)} directly by iteration, which reproduces
Eq.~\eqref{eq:(A11)} exactly after $N+1$~steps.

The gradient in~Eq.~\eqref{eq:(A10)} is obtained by introducing a ``probe''
field~$\chi$. This is a nilpotent variable (Grassmann-odd for a fermion
component) placed in a single field component~$j$ at the momentum~$-(r_I+cp)$,
and we treat it as an additional slot. We substitute
into~$\mathit{\Gamma}_{\text{int}}$ the ``available fields''
\begin{equation}
   \Phi_{\text{av}}
   =\Phi_{\text{bg}}+\sum_{\emptyset\neq I'\subseteq I}\delta\Phi_{I'}
   +\chi e_j,
\label{eq:(A13)}
\end{equation}
where $e_j$ is the unit vector of the component~$j$. Here, $\delta\Phi_{I'}$
are known for~$I'\subsetneq I$, and $\delta\Phi_I$ itself enters only through
the terms linear in it, i.e., $\sigma B\delta\Phi_I$
in~Eq.~\eqref{eq:(A10)}. Then, we have
\begin{equation}
   \left[\frac{\partial\mathit{\Gamma}_{\text{int}}}
   {\partial\Phi_j(-r_I-cp)}\right]_I
   =\text{coefficient of~$\chi$ in}~
   \left[\mathit{\Gamma}_{\text{int}}[\Phi_{\text{av}}]\right]_U,
   \qquad U:=I\cup\{\chi\},
\label{eq:(A14)}
\end{equation}
where $[\cdots]_U$ denotes the part that contains each element of~$U$ exactly
once. Since each term of~Eq.~\eqref{eq:(A13)} is a single Fourier mode with a
definite momentum, the convolutions corresponding to the local products
in~Eq.~\eqref{eq:(2.7)} reduce to products of the coefficients.

The functional~$\mathit{\Gamma}_{\text{int}}[\Phi_{\text{av}}]$ is constructed
directly from the definition of the ansatz~\eqref{eq:(2.7)}. The
$-1$~variables are built from~$\Phi_{\text{av}}$ as
\begin{align}
   \Psi_{-1}(p)
   &=\sum_{n\geq0}e_\tau^n\int_{k_1,\dots,k_n}
   w_n(p;k_1,\dots,k_n)_{\mu_1\cdots\mu_n}
\notag\\
   &\qquad{}
   \times
   e^{k_1^2}\mathcal{A}_{\mu_1}(k_1)\cdots e^{k_n^2}\mathcal{A}_{\mu_n}(k_n)
   e^{(p-\sum k_i)^2}\Psi(p-\sum_ik_i),
\label{eq:(A15)}
\end{align}
where the kernels~$w_n$, which carry no Dirac index, are obtained by solving
the flow equation~\eqref{eq:(2.10)} (see Appendix~B
of~Ref.~\cite{Nagao:2026gbq}), and $\Bar{\Psi}_{-1}$ is given
by~$e_\tau\to-e_\tau$. From these, we construct the kinetic term (with the free
part subtracted), the gauge vertex, and the four-Fermi terms. For the
four-Fermi terms, we decompose the
bilinears~$B_X=\Bar{\Psi}_{-1}\Gamma_X\Psi_{-1}$
as~$B_X=\sum_{V\subseteq U}B_X[V]$, where $B_X[V]$ is the part containing each
element of~$V$ exactly once. The relevant part
of~$-\frac{1}{2}g_X\int d^Dx\,B_X(x)^2$ in~Eq.~\eqref{eq:(A14)} is then
$-\frac{1}{2}g_X\sum_{V\subseteq U}B_X[V]{B_X[U\setminus V]}$, where
$(g_S,g_P,g_V,g_A)=(G_S,-G_S,G_V,G_V)$ and $\Gamma_S=1$, $\Gamma_P=\gamma_5$,
$\Gamma_V=\gamma_\mu$, and~$\Gamma_A=\gamma_5\gamma_\mu$. Note that, since
$\Psi_{-1}$ and~$\Bar{\Psi}_{-1}$ contain the gauge field to all orders, the
vertices of the type~$\Bar{\Psi}^2\Psi^2\mathcal{A}^n$ arising from the
four-Fermi terms are included automatically without being written down
explicitly, just as the vertices of the type~$\Bar{\Psi}\mathcal{A}^n\Psi$
arising from the kinetic term and the gauge vertex.

\subsection{Bookkeeping}
\label{sec:A.3}
In our numerical implementation, all quantities, such as~$\delta\Phi_I$
and~$\mathit{\Gamma}_{\text{int}}[\Phi_{\text{av}}]$, are kept exactly as polynomials
in the following variables: the Grassmann generators of the background fields,
i.e., the spinor components of~$\Bar{\Psi}_{\text{bg}}$ and~$\Psi_{\text{bg}}$
($\Bar{\Psi}_\alpha(0)$ and~$\Psi_\alpha(0)$ for the four-fermion part, and those
of~$\Bar{\Psi}_{\text{bg}}(-p)$ and~$\Psi_{\text{bg}}(p)$ for the fermion bilinear
part), the Grassmann generator~$\theta$ of the fermion slot, the probe~$\chi$,
the powers of~$e_\tau$, $G_S$, and~$G_V$, and ``tags'' that represent the powers
of the photon propagators at each sector momentum, $h_T((r_I+cp)^2)$
and~$h_L((r_I+cp)^2)$. With the tags, the photon part of~$J_0$
in~Eq.~\eqref{eq:(A7)} is kept symbolic, and the actual~$\mathcal{K}_\tau$
(and~$\xi_\tau$) are substituted at each step of the RG flow. Hence, the
multi-loop kernels, i.e., the integrands of the two- and three-loop
contributions in~Eq.~\eqref{eq:(A4)}, need to be tabulated only once as
functions of the loop momenta and of~$m_\tau$, which enters through~$h_F$.
For~$\xi_\tau=0$, the longitudinal part vanishes and only the three transverse
polarizations of the photon slots are needed; for general~$\xi_\tau$, all four
polarizations are summed. We have implemented these in a C++ program with
OpenMP parallelization.

\subsection{Projection}
\label{sec:A.4}
From the right-hand side thus obtained, we extract the following parts and
compare them with the left-hand side.

For the four-fermion part, we take the background fields at zero momentum and
expand this part as~$\sum_{\alpha<\beta,\,\gamma<\delta}c_{\alpha\beta\gamma\delta}
\Bar{\Psi}_\alpha(0)\Bar{\Psi}_\beta(0)\Psi_\gamma(0)\Psi_\delta(0)$, where
$6\times6=36$ monomials are independent because of the Grassmann nature. We
then project the $36$ coefficients~$c_{\alpha\beta\gamma\delta}$ by least squares onto
$O_S=(\Bar{\Psi}\Psi)^2-(\Bar{\Psi}\gamma_5\Psi)^2$,
$O_V=(\Bar{\Psi}\gamma_\mu\Psi)^2+(\Bar{\Psi}\gamma_5\gamma_\mu\Psi)^2$,
and~$O_X=(\Bar{\Psi}\Psi)^2+(\Bar{\Psi}\gamma_5\Psi)^2$, which form a
Fierz-complete basis of the rotation-invariant, parity-even structures
for~$N_f=1$; the least-squares projection is equivalent to the rotational
average. The coefficients of~$O_S$ and~$O_V$ give the loop contributions
to~$dG_{S\tau}/d\tau$ and~$dG_{V\tau}/d\tau$, which are added to the tree-level
term~$-(2+4\gamma_\psi)G_{Y\tau}$ ($Y=S$, $V$). The component along~$O_X$,
which is outside the ansatz, is discarded.

For the fermion bilinear part, we write the loop part of the coefficient
of~$\Bar{\Psi}_{\text{bg}}(-p)\cdots\Psi_{\text{bg}}(p)$
as~$\mathcal{R}(p)=a\Slash{p}+ib$. We obtain $a$ by the central difference
with~$p=\pm10^{-3}\Hat{e}_\lambda$ and the isotropic average
$a=\frac{1}{16}\sum_\lambda\partial_\lambda\tr(\gamma_\lambda\mathcal{R})$, and
$b$ from the trace. Then, $\gamma_\psi=a/2$ and~$dm_\tau/d\tau=(1-a)m_\tau+b$.

For the photon two-point function, i.e., the photon bilinear part, the one-loop
part and the two- and three-loop parts of~Eq.~(4.3)
of~Ref.~\cite{Nagao:2026vaf} are evaluated separately. As shown
in~Sect.~\ref{sec:A.1}, the four-Fermi interactions do not enter this part
directly.

\subsection{Loop integration}
\label{sec:A.5}
With the background fields at zero momentum, the one-loop contributions depend
on the slot momentum only through~$x=r_1^2$ and reduce to one-dimensional
integrals. We evaluate them on a grid of $221$~points ($x\in[0,36]$), on which
$\mathcal{K}_\tau(x)$ is represented by a cubic spline. For the two-loop
contributions ($C_2$, $Q_c$, $P_1$, $P_2$, and~$Q_b$ in~Eq.~\eqref{eq:(A4)}), we
set the slot momenta to~$r_1=|r_1|\Hat{e}_3$
and~$r_2=|r_2|(\sin\theta\Hat{e}_2+\cos\theta\Hat{e}_3)$, and use the
Gauss--Legendre quadrature with $16$~points for each of~$|r_1|$,
$|r_2|\in[0,3.6]$ and $10$~points for~$\theta$; the integrands decay
as~$e^{-2r_s^2}$. The three-loop contribution~$Q_a$, a $12$-dimensional integral
over the three slot momenta, is evaluated by scrambled Sobol sequences with a
Gaussian weight of width~$0.75$ ($512$--$1024$ samples). The multi-loop kernels
are tabulated at~$m_\tau=0$, $-0.3$, $-0.6$, $-0.9$, and~$-1.2$ and interpolated
by Lagrange polynomials on at most four nearest nodes.



%



\let\doi\relax









\begin{thebibliography}{99}

\bibitem{Maskawa:1974vs}
T.~Maskawa and H.~Nakajima,
Prog. Theor. Phys. \textbf{52}, 1326-1354 (1974)
doi:10.1143/PTP.52.1326

\bibitem{Maskawa:1975hx}
T.~Maskawa and H.~Nakajima,
Prog. Theor. Phys. \textbf{54}, 860 (1975)
doi:10.1143/PTP.54.860

\bibitem{Fukuda:1976zb}
R.~Fukuda and T.~Kugo,
Nucl. Phys. B \textbf{117}, 250-264 (1976)
doi:10.1016/0550-3213(76)90572-1

\bibitem{Nambu:1961tp}
Y.~Nambu and G.~Jona-Lasinio,
Phys. Rev. \textbf{122}, 345-358 (1961)
doi:10.1103/PhysRev.122.345

\bibitem{Nambu:1961fr}
Y.~Nambu and G.~Jona-Lasinio,
Phys. Rev. \textbf{124}, 246-254 (1961)
doi:10.1103/PhysRev.124.246

\bibitem{Kondo:1988qd}
K.~i.~Kondo, H.~Mino and K.~Yamawaki,
Phys. Rev. D \textbf{39}, 2430 (1989)
doi:10.1103/PhysRevD.39.2430

\bibitem{Miransky:1984ef}
V.~A.~Miransky,
Nuovo Cim. A \textbf{90}, 149-170 (1985)
doi:10.1007/BF02724229

\bibitem{Bardeen:1985sm}
W.~A.~Bardeen, C.~N.~Leung and S.~T.~Love,
Phys. Rev. Lett. \textbf{56}, 1230 (1986)
doi:10.1103/PhysRevLett.56.1230

\bibitem{Leung:1985sn}
C.~N.~Leung, S.~T.~Love and W.~A.~Bardeen,
Nucl. Phys. B \textbf{273}, 649-662 (1986)
doi:10.1016/0550-3213(86)90382-2

\bibitem{Holdom:1984sk}
B.~Holdom,
Phys. Lett. B \textbf{150}, 301-305 (1985)
doi:10.1016/0370-2693(85)91015-9

\bibitem{Yamawaki:1985zg}
K.~Yamawaki, M.~Bando and K.~i.~Matumoto,
Phys. Rev. Lett. \textbf{56}, 1335 (1986)
doi:10.1103/PhysRevLett.56.1335

\bibitem{Akiba:1985rr}
T.~Akiba and T.~Yanagida,
Phys. Lett. B \textbf{169}, 432-435 (1986)
doi:10.1016/0370-2693(86)90385-0

\bibitem{Appelquist:1986an}
T.~W.~Appelquist, D.~Karabali and L.~C.~R.~Wijewardhana,
Phys. Rev. Lett. \textbf{57}, 957 (1986)
doi:10.1103/PhysRevLett.57.957

\bibitem{Miransky:1988xi}
V.~A.~Miransky, M.~Tanabashi and K.~Yamawaki,
Phys. Lett. B \textbf{221}, 177-183 (1989)
doi:10.1016/0370-2693(89)91494-9

\bibitem{Miransky:1989ds}
V.~A.~Miransky, M.~Tanabashi and K.~Yamawaki,
Mod. Phys. Lett. A \textbf{4}, 1043 (1989)
doi:10.1142/S0217732389001210

\bibitem{Nambu:1989jt}
Y.~Nambu,
EFI-89-08.

\bibitem{Marciano:1989xd}
W.~J.~Marciano,
Phys. Rev. Lett. \textbf{62}, 2793-2796 (1989)
doi:10.1103/PhysRevLett.62.2793

\bibitem{Bardeen:1989ds}
W.~A.~Bardeen, C.~T.~Hill and M.~Lindner,
Phys. Rev. D \textbf{41}, 1647 (1990)
doi:10.1103/PhysRevD.41.1647

\bibitem{Hill:2002ap}
C.~T.~Hill and E.~H.~Simmons,
Phys. Rept. \textbf{381}, 235-402 (2003)
[erratum: Phys. Rept. \textbf{390}, 553-554 (2004)]
doi:10.1016/S0370-1573(03)00140-6
[arXiv:hep-ph/0203079 [hep-ph]].

\bibitem{Gockeler:1997dn}
M.~G\"ockeler, R.~Horsley, V.~Linke, P.~E.~L.~Rakow, G.~Schierholz and H.~St\"uben,
Phys. Rev. Lett. \textbf{80}, 4119-4122 (1998)
doi:10.1103/PhysRevLett.80.4119
[arXiv:hep-th/9712244 [hep-th]].

\bibitem{Wilson:1973jj}
K.~G.~Wilson and J.~B.~Kogut,
Phys. Rept. \textbf{12}, 75-199 (1974)
doi:10.1016/0370-1573(74)90023-4

\bibitem{Morris:1993qb}
T.~R.~Morris,
Int. J. Mod. Phys. A \textbf{9}, 2411-2450 (1994)
doi:10.1142/S0217751X94000972
[arXiv:hep-ph/9308265 [hep-ph]].

\bibitem{Becchi:1996an}
C.~Becchi,
[arXiv:hep-th/9607188 [hep-th]].

\bibitem{Pawlowski:2005xe}
J.~M.~Pawlowski,
Annals Phys. \textbf{322}, 2831-2915 (2007)
doi:10.1016/j.aop.2007.01.007
[arXiv:hep-th/0512261 [hep-th]].

\bibitem{Igarashi:2009tj}
Y.~Igarashi, K.~Itoh and H.~Sonoda,
Prog. Theor. Phys. Suppl. \textbf{181}, 1-166 (2010)
doi:10.1143/PTPS.181.1
[arXiv:0909.0327 [hep-th]].

\bibitem{Rosten:2010vm}
O.~J.~Rosten,
Phys. Rept. \textbf{511}, 177-272 (2012)
doi:10.1016/j.physrep.2011.12.003
[arXiv:1003.1366 [hep-th]].

\bibitem{Dupuis:2020fhh}
N.~Dupuis, L.~Canet, A.~Eichhorn, W.~Metzner, J.~M.~Pawlowski, M.~Tissier and N.~Wschebor,
Phys. Rept. \textbf{910}, 1-114 (2021)
doi:10.1016/j.physrep.2021.01.001
[arXiv:2006.04853 [cond-mat.stat-mech]].

\bibitem{Reuter:2019byg}
M.~Reuter and F.~Saueressig,
Cambridge University Press, 2019,
ISBN 978-1-107-10732-8, 978-1-108-67074-6

\bibitem{Aoki:1996fh}
K.~I.~Aoki, K.~i.~Morikawa, J.~I.~Sumi, H.~Terao and M.~Tomoyose,
Prog. Theor. Phys. \textbf{97}, 479-490 (1997)
doi:10.1143/PTP.97.479
[arXiv:hep-ph/9612459 [hep-ph]].

\bibitem{Kaplan:2009kr}
D.~B.~Kaplan, J.~W.~Lee, D.~T.~Son and M.~A.~Stephanov,
Phys. Rev. D \textbf{80}, 125005 (2009)
doi:10.1103/PhysRevD.80.125005
[arXiv:0905.4752 [hep-th]].

\bibitem{Gies:2004hy}
H.~Gies and J.~Jaeckel,
Phys. Rev. Lett. \textbf{93}, 110405 (2004)
doi:10.1103/PhysRevLett.93.110405
[arXiv:hep-ph/0405183 [hep-ph]].

\bibitem{Echigo:2025dia}
Y.~Echigo, Y.~Igarashi, K.~Itoh, J.~M.~Pawlowski and Y.~Takahashi,
PTEP \textbf{2026}, no.1, 013B01 (2026)
doi:10.1093/ptep/ptaf169
[arXiv:2507.12974 [hep-th]].

\bibitem{Sonoda:2020vut}
H.~Sonoda and H.~Suzuki,
PTEP \textbf{2021}, no.2, 023B05 (2021)
doi:10.1093/ptep/ptab006
[arXiv:2012.03568 [hep-th]].

\bibitem{Sonoda:2022fmk}
H.~Sonoda and H.~Suzuki,
PTEP \textbf{2022}, no.5, 053B01 (2022)
doi:10.1093/ptep/ptac047
[arXiv:2201.04448 [hep-th]].

\bibitem{Sonoda:2025tyu}
H.~Sonoda and H.~Suzuki,
PTEP \textbf{2025}, no.9, 093B05 (2025)
doi:10.1093/ptep/ptaf115
[arXiv:2506.23567 [hep-th]].

\bibitem{Sonoda:2019ibh}
H.~Sonoda and H.~Suzuki,
PTEP \textbf{2019}, no.3, 033B05 (2019)
doi:10.1093/ptep/ptz020
[arXiv:1901.05169 [hep-th]].

\bibitem{Miyakawa:2021hcx}
Y.~Miyakawa and H.~Suzuki,
PTEP \textbf{2021}, no.8, 083B04 (2021)
doi:10.1093/ptep/ptab100
[arXiv:2106.11142 [hep-th]].

\bibitem{Miyakawa:2021wus}
Y.~Miyakawa, H.~Sonoda and H.~Suzuki,
PTEP \textbf{2022}, no.2, 023B02 (2022)
doi:10.1093/ptep/ptac003
[arXiv:2111.15529 [hep-th]].

\bibitem{Abe:2022smm}
Y.~Abe, Y.~Hamada and J.~Haruna,
PTEP \textbf{2022}, no.3, 033B03 (2022)
doi:10.1093/ptep/ptac021
[arXiv:2201.04111 [hep-th]].

\bibitem{Miyakawa:2022qbz}
Y.~Miyakawa,
[arXiv:2201.08181 [hep-th]].

\bibitem{Miyakawa:2023yob}
Y.~Miyakawa, H.~Sonoda and H.~Suzuki,
PTEP \textbf{2023}, no.6, 063B03 (2023)
[erratum: PTEP \textbf{2024}, no.3, 039201 (2024)]
doi:10.1093/ptep/ptad074
[arXiv:2304.14753 [hep-th]].

\bibitem{Haruna:2023spq}
J.~Haruna and M.~Yamada,
JHEP \textbf{05}, 291 (2024)
doi:10.1007/JHEP05(2024)291
[arXiv:2312.15673 [hep-th]].

\bibitem{Nagao:2025sri}
S.~Nagao and H.~Suzuki,
PTEP \textbf{2025}, no.12, 123B03 (2025)
doi:10.1093/ptep/ptaf152
[arXiv:2508.16828 [hep-th]].

\bibitem{Narayanan:2006rf}
R.~Narayanan and H.~Neuberger,
JHEP \textbf{03}, 064 (2006)
doi:10.1088/1126-6708/2006/03/064
[arXiv:hep-th/0601210 [hep-th]].

\bibitem{Luscher:2010iy}
M.~L{\"u}scher,
JHEP \textbf{08}, 071 (2010)
[erratum: JHEP \textbf{03}, 092 (2014)]
doi:10.1007/JHEP08(2010)071
[arXiv:1006.4518 [hep-lat]].

\bibitem{Luscher:2011bx}
M.~L\"uscher and P.~Weisz,
JHEP \textbf{02}, 051 (2011)
doi:10.1007/JHEP02(2011)051
[arXiv:1101.0963 [hep-th]].

\bibitem{Luscher:2013cpa}
M.~L\"uscher,
JHEP \textbf{04}, 123 (2013)
doi:10.1007/JHEP04(2013)123
[arXiv:1302.5246 [hep-lat]].

\bibitem{Wilson:1979wp}
K.~G.~Wilson,
NATO Sci. Ser. B \textbf{59}, 363-402 (1980)
doi:10.1007/978-1-4684-7571-5{\_}20

\bibitem{Swendsen:1981rb}
R.~H.~Swendsen,
Phys. Rev. Lett. \textbf{47}, 1775-1777 (1981)
doi:10.1103/PhysRevLett.47.1775

\bibitem{Carosso:2018bmz}
A.~Carosso, A.~Hasenfratz and E.~T.~Neil,
Phys. Rev. Lett. \textbf{121}, no.20, 201601 (2018)
doi:10.1103/PhysRevLett.121.201601
[arXiv:1806.01385 [hep-lat]].

\bibitem{Nagao:2026vaf}
S.~Nagao and H.~Suzuki,
PTEP \textbf{2026}, no.7, 073B01 (2026)
doi:10.1093/ptep/ptag085
[arXiv:2603.11626 [hep-th]].

\bibitem{Nagao:2026gbq}
S.~Nagao and H.~Suzuki,
[arXiv:2610.19485 [hep-th]].

\bibitem{Nicoll:1977hi}
J.~F.~Nicoll and T.~S.~Chang,
Phys. Lett. A \textbf{62}, 287-289 (1977)
doi:10.1016/0375-9601(77)90417-0

\bibitem{Wetterich:1992yh}
C.~Wetterich,
Phys. Lett. B \textbf{301}, 90-94 (1993)
doi:10.1016/0370-2693(93)90726-X
[arXiv:1710.05815 [hep-th]].

\bibitem{Wetterich:1993ne}
C.~Wetterich,
Z. Phys. C \textbf{60}, 461-470 (1993)
doi:10.1007/BF01560044

\bibitem{Ginsparg:1981bj}
P.~H.~Ginsparg and K.~G.~Wilson,
Phys. Rev. D \textbf{25}, 2649 (1982)
doi:10.1103/PhysRevD.25.2649

\bibitem{Nagao:2026xxx}
S.~Nagao, H.~Sonoda and H.~Suzuki, in preparation.

\bibitem{Cresswell-Hogg:2026jpe}
C.~Cresswell-Hogg and D.~F.~Litim,
[arXiv:2603.19357 [hep-th]].

\bibitem{Bonini:1994kp}
M.~Bonini, M.~D'Attanasio and G.~Marchesini,
Nucl. Phys. B \textbf{437}, 163-186 (1995)
doi:10.1016/0550-3213(94)00569-Z
[arXiv:hep-th/9410138 [hep-th]].

\bibitem{Ellwanger:1995qf}
U.~Ellwanger, M.~Hirsch and A.~Weber,
Z. Phys. C \textbf{69}, 687-698 (1996)
doi:10.1007/s002880050073
[arXiv:hep-th/9506019 [hep-th]].

\bibitem{Gies:2020xuh}
H.~Gies and J.~Ziebell,
Eur. Phys. J. C \textbf{80}, 607 (2020)
doi:10.1140/epjc/s10052-020-8171-8
[arXiv:2005.07586 [hep-th]].

\bibitem{Gies:2022aiz}
H.~Gies, K.~K.~K.~Tam and J.~Ziebell,
Eur. Phys. J. C \textbf{83}, no.10, 955 (2023)
doi:10.1140/epjc/s10052-023-12080-y
[arXiv:2210.11927 [hep-th]].

\bibitem{Ellwanger:1996wy}
U.~Ellwanger, M.~Hirsch and A.~Weber,
Eur. Phys. J. C \textbf{1}, 563-578 (1998)
doi:10.1007/s100520050105
[arXiv:hep-ph/9606468 [hep-ph]].

\bibitem{Bergerhoff:1997cv}
B.~Bergerhoff and C.~Wetterich,
Phys. Rev. D \textbf{57}, 1591-1604 (1998)
doi:10.1103/PhysRevD.57.1591
[arXiv:hep-ph/9708425 [hep-ph]].

\bibitem{Mitter:2014wpa}
M.~Mitter, J.~M.~Pawlowski and N.~Strodthoff,
Phys. Rev. D \textbf{91}, 054035 (2015)
doi:10.1103/PhysRevD.91.054035
[arXiv:1411.7978 [hep-ph]].

\bibitem{Cyrol:2016tym}
A.~K.~Cyrol, L.~Fister, M.~Mitter, J.~M.~Pawlowski and N.~Strodthoff,
Phys. Rev. D \textbf{94}, no.5, 054005 (2016)
doi:10.1103/PhysRevD.94.054005
[arXiv:1605.01856 [hep-ph]].

\bibitem{Cyrol:2017ewj}
A.~K.~Cyrol, M.~Mitter, J.~M.~Pawlowski and N.~Strodthoff,
Phys. Rev. D \textbf{97}, no.5, 054006 (2018)
doi:10.1103/PhysRevD.97.054006
[arXiv:1706.06326 [hep-ph]].

\bibitem{Corell:2018yil}
L.~Corell, A.~K.~Cyrol, M.~Mitter, J.~M.~Pawlowski and N.~Strodthoff,
SciPost Phys. \textbf{5}, no.6, 066 (2018)
doi:10.21468/SciPostPhys.5.6.066
[arXiv:1803.10092 [hep-ph]].

\bibitem{Fu:2025hcm}
W.~j.~Fu, C.~Huang, J.~M.~Pawlowski, Y.~y.~Tan and L.~j.~Zhou,
Phys. Rev. D \textbf{112}, no.5, 054047 (2025)
doi:10.1103/4sh5-w4yc
[arXiv:2502.14388 [hep-ph]].



\end{thebibliography}


\end{document}